\documentclass[preprint,12pt,sort,compress]{elsarticle}
\usepackage{amssymb}
\usepackage{amsthm}
\usepackage{amsmath}
\usepackage{mathtools}
\usepackage{mathrsfs}
\usepackage[algo2e]{algorithm2e}
\usepackage{array}
\usepackage{multirow}
\usepackage{listings}
\usepackage{tabu}
\usepackage{booktabs}
\usepackage{enumerate}
\usepackage{fullpage}
\usepackage{float}
\usepackage{xcolor}
\usepackage[colorinlistoftodos]{todonotes}
\usepackage{bbm}
\usepackage{bm}
\usepackage[colorlinks=true]{hyperref}
\usepackage{url}
\usepackage{textcomp}
\usepackage{gensymb}
\usepackage{lineno}
\usepackage{graphicx}
\usepackage{float}
\usepackage{caption}
\usepackage{tikz}
\usetikzlibrary{shapes}
\usepackage{graphicx} % Required for inserting images
\usepackage{multirow}
\usepackage{overpic}
\usepackage{subcaption}
\usepackage{enumitem}
\usepackage{footnote}
\usepackage{parskip}
\usepackage{pifont} % For nice ticks and crosses
\usepackage{adjustbox}
\usepackage{booktabs, multirow} % for borders and merged ranges
\usepackage{soul}% for underlines
\usepackage{xcolor,colortbl} % for cell colors
\usepackage{tikz}
\usepackage{cleveref}
\usetikzlibrary{calc}
\usepackage[table]{xcolor} % needed for cell coloring
\usepackage{multirow, booktabs}
\definecolor{cpink}{HTML}{FCCDE5}
\definecolor{cred}{HTML}{FFC2BA}
\definecolor{cyellow}{HTML}{FFFFB3}
\definecolor{cblue}{HTML}{B9DEFF}
\definecolor{cgreen}{HTML}{D7F3E7}
\definecolor{cneutral}{HTML}{CFCFCF}

\author[cornellMAE]{Junyi Guo \fnref{equal}}
\author[ndAME]{Pan Du \fnref{equal}}
\author[cornellMAE,ndAME]{Xiantao Fan}
\author[ndAME]{Yahui Li}
\author[cornellMAE,ndAME]{Jian-Xun Wang\corref{corxh}}

\address[cornellMAE]{Sibley School of Mechanical and Aerospace Engineering, Cornell University, Ithaca, NY, USA}
\address[ndAME]{Department of Aerospace and Mechanical Engineering, University of Notre Dame, Notre Dame, IN}

\cortext[corxh]{Corresponding author. Tel: +1 540 3156512}
\ead{jw2837@cornell.edu}
\fntext[equal]{Equal contribution}

\newcommand{\Dcal}{\mathcal{D}}
\newcommand{\Tcal}{\mathcal{T}}
\newcommand{\qvb}{\mathbf{q}}

\newcommand{\R}{\mathbb{R}}

\biboptions{numbers,comma,round,square}
\journal{Elsevier}

\begin{document}
\begin{frontmatter}
\title{Conditional neural field for spatial dimension reduction of turbulence data: a comparison study}

\begin{abstract}
We investigate {conditional neural fields} (CNFs), mesh-agnostic, coordinate-based decoders conditioned on a low-dimensional latent, for spatial dimensionality reduction of turbulent flows. CNFs are benchmarked against Proper Orthogonal Decomposition and a convolutional autoencoder within a unified encoding–decoding framework and a common evaluation protocol that explicitly separates {in-range} (interpolative) from {out-of-range} (strict extrapolative) testing beyond the training horizon, with identical preprocessing, metrics, and fixed splits across all baselines. We examine three conditioning mechanisms: (i) activation-only modulation (often termed FiLM), (ii) low-rank weight + bias modulation (termed FP), and (iii) last-layer inner-product coupling, and introduce a novel domain-decomposed CNF that localizes complexities. Across representative turbulence datasets (WMLES channel inflow, DNS channel inflow, and wall pressure fluctuations over turbulent boundary layers), CNF–FP achieves the lowest training and in-range testing errors, while CNF–FiLM generalizes best for out-of-range scenarios once moderate latent capacity is available. Domain decomposition significantly improves out-of-range accuracy, especially for the more demanding datasets.. The study provides a rigorous, physics-aware basis for selecting conditioning, capacity, and domain decomposition when using CNFs for turbulence compression and reconstruction.

\end{abstract}

\end{frontmatter}

\section{Introduction}
\label{sec:intro}
%Check makrs

Turbulent flows are characterized by high-dimensional, multi-scale spatiotemporal structures that pose significant challenges in computational fluid dynamics (CFD), both from computational and storage perspectives \cite{pope2001turbulent}. The detailed analysis, visualization, and interpretation of turbulence data obtained from high-fidelity numerical simulations, such as direct numerical simulation (DNS) or large-eddy simulation (LES), typically demand substantial computational resources. Consequently, there is a strong motivation to represent such high-dimensional data efficiently by encoding turbulent fields into compact, low-dimensional latent embeddings. These latent representations are valuable not only for flow modal analysis~\cite{taira2017modal,taira2020modal}, but also for downstream tasks such as reduced-order modeling~\cite{kim2022fast,khoo2022sparse,solera2024beta,fresca2022pod}, surrogate modeling~\cite{ren2021phycrnet,sun2020surrogate,sun2023unifying}, flow reconstruction~\cite{ren2023physr,Erichson_2020,fukami2021machineSR}, and optimization~\cite{zhang2024novel}, all of which often require accurate inversion from latent space back to the physical domain. 

Classical linear dimensionality reduction methods, notably Proper Orthogonal Decomposition (POD)~\cite{lumley1967structure,picard2000pressure} and Dynamic Mode Decomposition (DMD)~\cite{schmid2010dynamic}, provide well-established frameworks for projecting turbulent fields onto low-dimensional linear subspaces. POD identifies modes capturing maximal energy content, while DMD extracts coherent structures associated with characteristic frequencies and growth rates. Despite their utility and interpretability, these linear methods inherently struggle to represent strongly nonlinear features and multiscale structures of turbulence, often necessitating a prohibitively large number of modes to achieve acceptable accuracy~\cite{csala2022comparing,mendez2023linear}.

To better address nonlinear and complex turbulent structures, recent studies have increasingly explored nonlinear dimensionality reduction (NDR) techniques, offering a compelling alternative by discovering manifolds that better conform to the data's intrinsic geometry. In fields like computer vision, techniques such as Kernel Principal Component Analysis (KPCA)~\cite{scholkopf1998nonlinear}, isometric mapping (Isomap)~\cite{balasubramanian2002isomap}, Locally Linear Embedding (LLE)~\cite{roweis2000nonlinear}, and t-distributed stochastic neighbor embedding (t-SNE)~\cite{van2008visualizing} have shown success in unfolding nonlinear manifolds~\cite{lee2007nonlinear}. In fluid dynamics, however, these manifold learning techniques have seen limited application until recently. One significant challenge is that many NDR methods lack a straightforward inverse mapping from latent space back to physical space - commonly known as the pre-image problem~\cite{kwok2004pre}. Without a reliable way to reconstruct the flow field from the reduced coordinates, these techniques are of limited use for compression or surrogate modeling. For example, KPCA can embed data nonlinearly, but computing an approximate inverse is non-trivial and often requires solving additional nonconvex optimization problems. Similarly, techniques like Isomap or t-SNE are primarily geared toward visualization or clustering, and do not provide an explicit decoder to generate flow fields from latents.

Recent advances in deep learning (DL) have introduced neural-network-based autoencoders, especially convolutional neural network autoencoders (CNN-AEs), which simultaneously learn nonlinear encoding and decoding mappings for effective compression and accurate reconstruction of turbulent fields~\cite{murata2020nonlinear,Erichson_2020,Fukami_2021}. For instance, Murata et al.\cite{murata2020nonlinear} demonstrated that a CNN-AE significantly outperformed POD in reconstructing cylinder wake flows, achieving substantially lower errors for an equivalent latent dimensionality. CNN-AEs have further demonstrated efficacy in reconstructing high-fidelity fields from coarse simulations~\cite{Erichson_2020,ren2023physr} and in accelerating fluid dynamic simulations via latent-space dynamics forecasting~\cite{Fukami_2021,Racca_Doan_Magri_2023}. By leveraging convolutional layers, CNN-AEs efficiently capture spatial correlations inherent in data, thereby providing enhanced reconstruction fidelity compared to traditional linear methods. Nonetheless, CNN-AEs typically rely on structured grid data, which restricts their applicability to irregular, unstructured, or adaptively meshed flow domains frequently encountered in practical CFD applications.

A promising alternative is the emerging paradigm of neural field (NF) representations, coordinate-based neural networks that parameterize flow fields as continuous, implicit functions of spatial coordinates. Such representations are mesh-agnostic and naturally handle unstructured or adaptive grids – one can query field values at any coordinate, irrespective of how the training data were sampled~\cite{xie2022neural}. They also provide implicit continuous resolution: the network output can be evaluated on a finer grid than it was trained on, enabling super-resolution of the field without an explicit interpolation steps~\cite{gao2022nerf}. These features make NFs especially attractive for fluid dynamics, where geometric flexibility and multiscale resolution are often required. A conditional neural field (CNF) extends this concept to represent not just one fixed field but an entire family of fields conditioned on a latent code. In practice, a CNF is realized by augmenting the input of the NF network with a latent vector $\mathbf{z}$ that encodes the identity of a particular flow snapshot or flow condition. The latent $\mathbf{z}$ plays a role analogous to the code in an AE's decoder – it is a compact description of the specific flow instance. The mapping from latent to physical space is deterministic and invertible in the sense that, given the trained CNFs, each $\mathbf{z}$ produces a unique field. Recovering $\mathbf{z}$ from a new field requires either an encoder network or optimization (i.e., auto-decoding)~\cite{bojanowski2018optimizing,park2019deepsdf}. This framework thus meets the criterion of invertible encoding by design. 

Crucially, CNFs retain the advantages of implicit neural representations (INRs), as they do not require training data on a fixed mesh; one can train on a variety of discretizations or even point cloud samples of the field. Chen et al.~\cite{chen2023crom,chen2023implicitneuralspatialrepresentations} highlight this in their continuous reduced-order modeling (CROM) approach: rather than building a basis for a fixed grid of PDE solutions, they construct a low-dimensional embedding of the continuous vector fields themselves using CNFs, enabling training on data from diverse grids and producing a single latent space characterizing the continuous solution manifold. Serrano et al.~\cite{serrano2023operator} further demonstrate the versatility of CNFs for operator learning in PDEs, showing their effectiveness in tackling PDEs defined on general geometries and highlighting their potential in broader scientific computing applications. Similarly, Yin et al.~\cite{yin2022continuous} leveraged these implicit neural representations to accurately forecast continuous PDE dynamics, effectively capturing both low-frequency and high-frequency content. Another strength of NFs, particularly relevant for turbulence, is the inherent capability to incorporate multi-scale details. Standard multilayer perceptrons with smooth activation functions exhibit a known spectral bias, prioritizing low-frequency (smooth) components and struggling with high-frequency content~\cite{rahaman2019spectral}. Turbulent fields, with their eddies and sharp gradients, inherently contain a wide range of frequencies. To address this, recent advances like periodic activation functions (SIREN networks)~\cite{sitzmann2020implicit} and Fourier feature embeddings~\cite{tancik2020fourier} can be employed to better represent fine-scale structures. Pan et al.~\cite{pan2023neural} demonstrated the efficacy of CNFs specifically for spatial dimension reduction and reconstruction of three-dimensional turbulent flows, highlighting their superior performance compared to classical linear methods. More recently, Du et al.~\cite{du2024confild} introduced CoNFiLD, integrating latent diffusion models with CNFs to generate realistic spatiotemporal turbulence fields conditioned on partial or sparse observations, which has been successfully applied to inflow turbulence generation~\cite{liu2025confild}. 

Despite these promising developments, CNFs remain relatively unexplored in turbulence modeling. They have yet to be systematically benchmarked against classical linear dimension reduction methods, such as POD, or widely-used nonlinear DL-based approaches, e.g., CNN-AEs. More importantly, generalizability, particularly extrapolation to flow conditions or time horizons beyond the training regime, has rarely been explicitly tested for most DL-based NDR models. Previous studies typically assess reconstruction accuracy primarily within training datasets or through interpolation over narrow parameter ranges~\cite{csala2022comparing,zhang2023comparison,pan2023neural}. However, real-world applications often demand robust performance under scenarios far outside the original training distribution. Another unresolved question pertains to how the conditioning mechanism (i.e., the way the latent code is incorporated into neural representations) impacts model performance for turbulent flows. While advanced conditioning strategies such as feature-wise linear modulation (FiLM) have demonstrated improved efficacy in certain contexts~\cite{du2024confild}, their suitability and effectiveness specifically for turbulence representation remain underexplored. To address these gaps, we propose a unified framework for systematically comparing CNFs with classical linear and nonlinear ML-based dimension reduction approaches, enabling a consistent evaluation of reconstruction accuracy and generalization capabilities across diverse turbulence datasets. Furthermore, we introduce a novel domain-decomposition strategy within the CNF framework specifically tailored to handle large-scale, highly complex turbulent flow data, aiming to significantly enhance reconstruction accuracy and improve generalization. By explicitly testing interpolation and extrapolation performance beyond the training horizons using both quantitative reconstruction metrics and qualitative assessments of physical fidelity, we investigate how structured latent-space architectures and domain decomposition can strengthen model robustness. Ultimately, this work seeks to establish CNFs, combined with our proposed improvements, as accurate, reliable, and practically viable NDR tools for turbulence data compression, reconstruction, and analysis.
 
The remainder of this paper is structured as follows: In section \ref{sec:method}, we introduce CNFs for NDR with different conditioning mechanisms and formulate all baseline dimension reduction methods within a unified framework. In Section~\ref{sec:Result}, we report the benchmarking results and explicitly compare the extrapolation and interpolation performance of each method. Finally, Section~\ref{sec:Discussion} discusses broader implications of our findings and outlines potential directions for future research.

\section{Methodology}
\label{sec:method}

In this section, we present a unified framework for spatial dimension reduction and reconstruction of turbulent flow fields. We systematically introduce and compare classical POD, CNN-AEs, and CNF-based dimension reduction methods.

\subsection{Unified framework for spatial dimension reduction}

We introduce a generalized encoding-decoding framework for spatial dimension reduction of high-dimensional flow fields, structured to systematically represent a broad class of linear and nonlinear methods. This generic framework provides a clear mathematical foundation that covers the fundamental principles shared across diverse dimension reduction techniques, facilitating their comparative analysis and consistent evaluation.

In general, consider a spatiotemporal scalar field $q(\mathbf{x}, t)$, discretized in space and time forming the snapshot matrix $\boldsymbol{\Phi} = \big[ \boldsymbol{\phi}^1, \, \boldsymbol{\phi}^2, \, \ldots, \, \boldsymbol{\phi}^n \big] = \big[ q(t_1), \, q(t_2), \, \ldots, \, q(t_n) \big] \in \mathbb{R}^{m \times n}$, where $m$ is the spatial dimension (e.g., the total number of grid points), and $n$ denotes the number of temporal snapshots. The objective of spatial dimension reduction techniques is to identify a compact, low-dimensional latent representation $\mathbf{Z} = \big[ \mathbf{z}^1, \, \mathbf{z}^2, \, \ldots, \, \mathbf{z}^n \big] \in \mathbb{R}^{r \times n}$ that effectively captures the primary features of the original high-dimensional fields, facilitating efficient storage, analysis, and reconstruction.

The dimension reduction process can be defined through two core mathematical operations: (1) an \emph{encoding} operation, mapping the original field into a reduced latent space, and (2) a \emph{decoding} operation, mapping the latent variables back into the original spatial representation.

\paragraph{Encoding operation}
Let $\mathcal{E}$ denote a generic linear or nonlinear encoding operator parameterized by a set of parameters $\boldsymbol{\theta}^e$. This operator compresses each snapshot in the original spatiotemporal data $\boldsymbol{\Phi}$ into its lower-dimensional latent representation $\mathbf{Z}$, defined as:
\begin{equation}\label{eq:encoder_general}
\mathbf{z}^i = \mathcal{E}(\boldsymbol{\phi}^i; \boldsymbol{\theta}^e), 
\quad \boldsymbol{\phi}^i \in \mathbb{R}^m, \;
\mathbf{z}^i \in \mathbb{R}^r, \;
i = 1,2,\ldots,n, \;
r \ll m,
\end{equation}

where the latent representation $\mathbf{z}^i$ encodes the essential spatial structures of the flow fields into a significantly reduced-dimensional form. We uniquely interpret the encoding process $\mathcal{E}$ as a two-step procedure. The first step, a transformation, maps the high-dimensional snapshot into a feature space that reorganizes the representation without loss of information, making it more amenable to compression. The second step, a reduction, projects this transformed representation onto a compact latent space, thereby discarding redundancy and retaining only the most essential flow structures.

\begin{itemize}
\item \emph{Transformation step} ($\mathcal{T}^e$): This step maps the $i^{th}$ data snapshot $\boldsymbol{\phi}^i$ into an intermediate feature representation $\boldsymbol{\gamma}^i$:
\begin{equation}\label{eq:transformation_general}
\boldsymbol{\gamma}^i = \mathcal{T}^e(\boldsymbol{\phi}^i; \boldsymbol{\theta}^e_T), \quad \boldsymbol{\gamma}^i \in \mathbb{R}^{s},
\end{equation}
where $\boldsymbol{\gamma}^i \in \mathbb{R}^{s}$ is an intermediate representation. Depending on the chosen method, the transformation may correspond, among others, to a linear projection (e.g., POD or Fourier bases), a nonlinear mapping (e.g., kernel embeddings), or a learned operator such as convolutional layers that capture localized flow features.

\item \emph{Reduction step} ($\mathcal{R}^e$): The subsequent reduction step explicitly projects the intermediate representation $\boldsymbol{\gamma}^i$ onto the lower-dimensional latent space:
\begin{equation}\label{eq:reduction_general}
    \mathbf{z}^i = \mathcal{R}^e(\boldsymbol{\gamma}^i; \boldsymbol{\theta}^e_R), \quad  \mathbf{z}^i  \in \mathbb{R}^{r}. 
\end{equation}
Here, $\mathcal{R}^e$ may be realized through a variety of approaches, ranging from simple pooling or averaging operations, to rank-reducing linear transformations, to learned nonlinear projections implemented by neural networks.
\end{itemize}
The full encoding operation can be written as the composition of the transformation and reduction steps:
\begin{equation}\label{eq:full_encoding_composite}
\mathbf{z}^i = \mathcal{R}^e_{\boldsymbol{\theta}^e_R} \circ \mathcal{T}^e_{\boldsymbol{\theta}^e_T}(\boldsymbol{\phi}^i) = \mathcal{E}_{\boldsymbol{\theta}^e}(\boldsymbol{\phi}^i), \quad \boldsymbol{\theta}^e = \{\boldsymbol{\theta}^e_{T}, \boldsymbol{\theta}^e_{R}\}
\end{equation}

\paragraph{Decoding operation}
The decoding operation defines an inverse mapping from the low-dimensional latent space $\mathbf{z}^i $ back to the original high-dimensional space. Formally, the decoding operator $\mathcal{D}$, parameterized by decoder parameters $\boldsymbol{\theta}^d$, is defined as:
\begin{equation}\label{eq:decoder_general}
\hat{\boldsymbol{\phi}}^i = \mathcal{D}(\mathbf{z}^i; \boldsymbol{\theta}^d), 
\quad \hat{\boldsymbol{\phi}}^i \in \mathbb{R}^{m}, 
\quad i = 1,2,\ldots,n,
\end{equation}
which produces reconstructed fields $\hat{\boldsymbol{\phi}}^i$. Analogous to the encoder, the decoder $\mathcal{D}$ can also be decomposed into two counterpart sub-operators:  
\begin{itemize}
\item \emph{Reverse reduction step} ($\mathcal{R}^d$): 
This step maps the latent code $\mathbf{z}^i$ back into the intermediate representation space:
\begin{equation}\label{eq:inverse_reduction}
\hat{\boldsymbol{\gamma}}^i = \mathcal{R}^d(\mathbf{z}^i; \boldsymbol{\theta}^d_R), 
\quad \hat{\boldsymbol{\gamma}}^i \in \mathbb{R}^s.
\end{equation}

\item \emph{Reverse transformation step} ($\mathcal{T}^{d}$): 
Subsequently, this step maps the intermediate representation $\hat{\boldsymbol{\gamma}}^i$ 
back to the original high-dimensional snapshot:
\begin{equation}\label{eq:inverse_transformation}
\hat{\boldsymbol{\phi}}^i = \mathcal{T}^d(\hat{\boldsymbol{\gamma}}^i; \boldsymbol{\theta}^d_T), 
\quad \hat{\boldsymbol{\phi}}^i \in \mathbb{R}^m.
\end{equation}
\end{itemize}
Thus, the complete decoding process can be expressed explicitly as:
\begin{equation}
\label{eq:full_decoding_composite}
\hat{\boldsymbol{\phi}}^i =\mathcal{T}^d_{\boldsymbol{\theta}^d_T} \circ \mathcal{R}^d_{\boldsymbol{\theta}^d_R}(\mathbf{z}^i)
\end{equation}

\paragraph{Optimization}
Given training data samples drawn from a distribution $\mathcal{G}$, the dimension reduction model parameters $\{\boldsymbol{\theta}^e_T, \boldsymbol{\theta}^e_R, \boldsymbol{\theta}^d_T, \boldsymbol{\theta}^d_R\}$ are identified through an optimization problem that seeks to minimize the reconstruction error between original and reconstructed fields:
\begin{equation}
\label{eq:optimization_general}
\min_{\substack{\boldsymbol{\theta}_T^e,\, \boldsymbol{\theta}_R^e, \boldsymbol{\theta}_T^d,\, \boldsymbol{\theta}_R^d}} 
\; \mathbb{E}_{\boldsymbol{\phi}\sim \mathcal{G}}\!\left[ 
\sum_{i=1}^n \mathcal{L}\left(\boldsymbol{\phi}^i, 
\mathcal{D}_{\boldsymbol{\theta}^d}\circ \mathcal{E}_{\boldsymbol{\theta}^e}(\boldsymbol{\phi}^i)\right)
\right],
\end{equation}
where $\mathcal{L}$ denotes a loss function that measures the discrepancy between the original snapshot and its reconstruction.
Depending on the chosen method, $\mathcal{T}^e$, $\mathcal{R}^e$, and their decoder counterparts may represent linear transformations (e.g., singular value decomposition in POD), nonlinear mappings (e.g., kernel-based embeddings), or parameterized neural network layers (e.g., convolutional or fully connected layers in deep learning). This formulation thus provides a unifying mathematical framework that encompasses a wide range of dimensionality reduction methodologies, including classical POD, convolutional autoencoders (CNN-AEs), and variants of CNFs.
\begin{figure}[t!]
    \centering
    \includegraphics[width=1\linewidth]{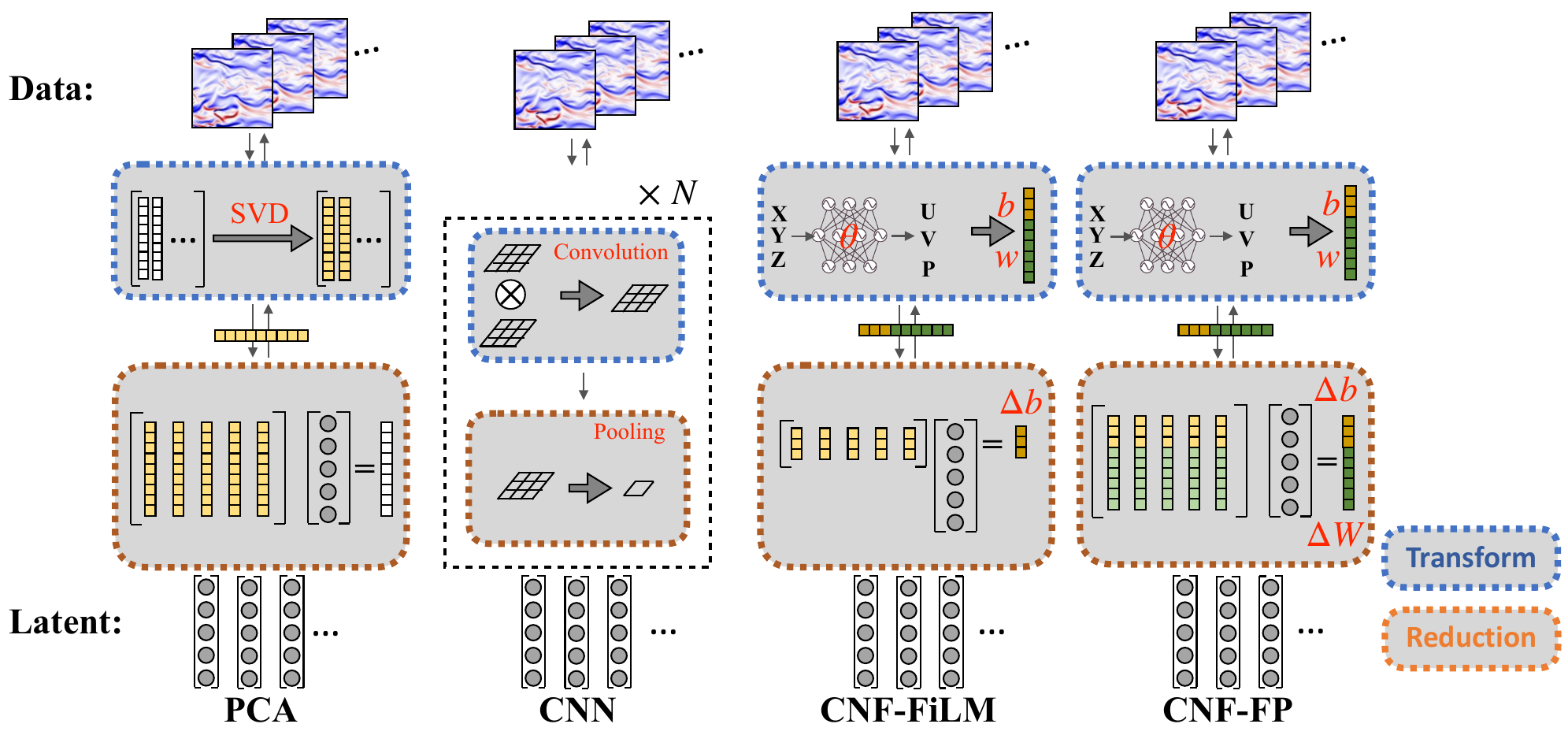}
    \caption{Schematic illustration of the unified encoding-decoding framework for spatial dimension reduction methods, demonstrating the common structure comprising transformation (blue dashed boxes) and reduction (orange dashed boxes) steps. Representative methods shown include linear (PCA/POD), convolutional neural network autoencoder (CNN-AE), and conditional neural fields (CNF-FiLM and CNF-FP)}
    \label{fig:method_overview}
\end{figure}
As illustrated in Figure~\ref{fig:method_overview}, these methods can all be represented in terms of a transformation step (blue dashed boxes) and a reduction step (orange dashed boxes). This perspective highlights their shared conceptual foundations and enables direct comparison of their respective capabilities in dimensionality reduction and reconstruction fidelity. In the following subsections, each method will be formally introduced and analyzed within this unified encoding–decoding framework.

\subsection{Baseline dimension reduction methods}

\subsubsection{Proper orthogonal decomposition (POD)}
POD is classically defined via the singular value decomposition (SVD) of the snapshot matrix $\boldsymbol{\Phi} \in \R^{m\times n}$~\cite{eckart1936approximation}:
\begin{equation}
    \boldsymbol{\Phi}= \mathbf{U} \boldsymbol{\Sigma} \mathbf{V}^T, 
\end{equation}
%where columns of $\mathbf{U} \in \R^{m \times n}$ are spatial orthonormal modes, $\boldsymbol{\Sigma} \in \R^{n \times n}$ a diagonal matrix representing mode energy, columns of $\mathbf{V} \in \R^{n \times n}$ are mode coefficients for each snapshot. 
where $\mathbf{U}\in\mathbb{R}^{m\times n}$ contains spatial orthonormal modes (columns),
$\boldsymbol{\Sigma}\in\mathbb{R}^{n\times n}$ holds singular values, and
$\mathbf{V}\in\mathbb{R}^{n\times n}$ contains temporal right singular vectors. 
The POD coefficients for snapshot $i$ are $\mathbf{a}^i=\boldsymbol{\Sigma}\,\mathbf{V}^{\top}\mathbf{e}_i$
(equivalently $\mathbf{a}^i=\mathbf{U}^{\top}\boldsymbol{\phi}^i$).
Reinterpreting using our proposed framework, we can rewrite the encoding-decoding processes as follows: 
\paragraph{Encoding operation} 
The encoding operation for POD is explicitly defined as,
\begin{itemize}
\item \emph{Transformation step} ($\mathcal{T}$): The original snapshot $\boldsymbol{\phi}^i$ is linearly projected onto the complete spatial POD basis $\mathbf{U}$, yielding an intermediate representation,
\begin{equation}
    \boldsymbol{\gamma}^i = \mathcal{T}^e(\boldsymbol{\phi}^i; \boldsymbol{\theta}^e_T) = \mathbf{U}^T \boldsymbol{\phi}^i,
\end{equation}
where the parameter $\boldsymbol{\theta}^e_T$ is the POD spatial basis $\mathbf{U}$ itself.

\item \emph{Reduction step} ($\mathcal{R}$): 
The subsequent reduction step truncates the basis to the leading $r$ dominant components of the intermediate representation $\boldsymbol{\gamma}^i$, yielding the reduced-dimensional latent vector:

\begin{equation}
    \mathbf{z}^i = \mathcal{R}^e(\boldsymbol{\gamma}^i; \boldsymbol{\theta}^e_R) = \boldsymbol{\gamma}^i[:r],
    \quad \mathbf{z}^i \in \mathbb{R}^{r},
    %\bm{z}(t) = \mathcal{R}(\bm{\gamma}(t); \boldsymbol{\theta}_R) = \bm{U}_r^T\bm{\phi}(t),
\end{equation}
where $\boldsymbol{\gamma}^i[:r]$ denotes selecting the first $r$ components of $\boldsymbol{\gamma}^i$, corresponding to the dominant energetic features of the original flow data. In this classical linear setting, no parameters are involved in the reduction step (i.e., $\boldsymbol{\theta}_R^e = \emptyset$).

\end{itemize}

\paragraph{Decoding operation}
The decoding operation reconstructs the snapshot by linearly combining the retained orthonormal spatial modes with the latent coefficients $\boldsymbol{z}(t)$.

\begin{itemize}
\item \emph{Reverse reduction step} ($\mathcal{R}^{d}$):
Given the latent vector $\mathbf{z}^i$, the intermediate representation is reconstructed by padding zeros to restore the original dimension $n$:
\begin{equation}
    \hat{\boldsymbol{\gamma}}^i = \mathcal{R}^{d}(\mathbf{z}^i) =
\begin{bmatrix}
\mathbf{z}^i \ 
\mathbf{0}
\end{bmatrix}, \quad \hat{\mathbf{\gamma}}^i\in \mathbb{R}^{n},
\end{equation}
where zeros are padded to the latent vector to restore its dimension to match the full set of spatial modes.
\item \emph{Reverse transformation step} ($\mathcal{T}^{d}$):
The reconstructed high-dimensional field is obtained by linearly combining the spatial POD modes
\begin{equation}
    \hat{\boldsymbol{\phi}}^{i} = \mathcal{T}^{d}(\hat{\boldsymbol{\gamma}}^i; \boldsymbol{\theta}^d_T) 
    = \mathbf{U}\hat{\boldsymbol{\gamma}}^i, 
    \quad \hat{\boldsymbol{\phi}}^i\in \mathbb{R}^{m},
\end{equation}
where $\boldsymbol{\theta}^d_T = \boldsymbol{\theta}^e_T$ is also the POD basis $\mathbf{U}$.

\end{itemize}
\paragraph{Optimization problem}
In the context of POD, the optimization problem seeks an $r$-dimensional basis that minimizes the projection error, measured in the squared Frobenius norm of the snapshot matrix:
\begin{equation}
\min_{\bm{U}_r}\|\bm{\Phi}-\bm{U}_r\bm{U}_r^{T}\bm{\Phi}\|^2_{F},
\end{equation}
whose solution, by the Eckart–Young–Mirsky theorem~\cite{chu2003structured}, is given by the truncated SVD of $\boldsymbol{\Phi}$. 
This formulation allows POD to be reinterpreted within the unified encoding–decoding framework, facilitating direct comparison with modern deep learning–based dimensionality reduction approaches such as CNN-AEs and CNFs.

\subsubsection{Convolutional autoencoder (CNN-AE)}
The CNN-AE performs spatial dimensionality reduction and reconstruction progressively. At each stage, \emph{transformation} (convolutional message passing) and \emph{reduction} (stride/pooling) are interleaved; Likewise, the decoding process interleaves reverse reduction (upsampling) and reverse trans-
formation (feature reconstruction) steps. we model this with \emph{micro-blocks} that fit cleanly into the unified framework.

\paragraph{Encoding operation}
Let $H^{(0)}=\mathrm{reshape}(\boldsymbol{\phi}^i)$ be the input feature map. For $\ell=1,\ldots,L$ we apply a micro-block
\begin{equation}
\tilde{H}^{(\ell)}=\underbrace{\mathcal{T}^{e,\ell}\!\big(H^{(\ell-1)};\boldsymbol{\theta}^{e,\ell}_T\big)}_{\text{message passing: conv/BN/activation}},\qquad
H^{(\ell)}=\underbrace{\mathcal{R}^{e,\ell}\!\big(\tilde{H}^{(\ell)}\big)}_{\text{reduction: stride or pooling}},
\end{equation}
and denote $\mathcal{B}_\ell:=\mathcal{R}^{e,\ell}\circ \mathcal{T}^{e,\ell}$. After $L$ blocks we obtain an intermediate vector
\begin{equation}
\boldsymbol{\gamma}^i=\mathrm{vec}\!\big(\mathcal{B}_L\circ\cdots\circ\mathcal{B}_1(\boldsymbol{\phi}^i)\big)\in\mathbb{R}^{s},
\end{equation}
which is projected to the latent space by a bottleneck map:
\begin{equation}
\mathbf{z}^i=\mathcal{R}^{e,\mathrm{bottleneck}}\!\big(\boldsymbol{\gamma}^i;\boldsymbol{\theta}^{e}_R\big)\in\mathbb{R}^{r}.
\end{equation}
\emph{Remark.} While downsampling reduces spatial resolution locally at each stage, we treat these as \emph{micro-reductions} inside the hierarchical $\mathcal{T}^e$ pathway and reserve $\mathcal{R}^{e,\mathrm{bottleneck}}$ for the final projection to dimension $r$.

\paragraph{Decoding operation}
Starting from the latent, we first expand back to the encoder's terminal feature size and reshape:
\begin{equation}
\hat{\boldsymbol{\gamma}}^i=\mathcal{R}^{d,\mathrm{expand}}\!\big(\mathbf{z}^i;\boldsymbol{\theta}^{d}_R\big)\in\mathbb{R}^{s}
\ \xrightarrow{\ \mathrm{reshape}\ }\ \hat{H}^{(L)}.
\end{equation}
Then for $\ell=L,\ldots,1$ we mirror the micro-blocks with upsampling (or transposed stride) followed by convolution:
\begin{equation}
\tilde{H}^{(\ell-1)}=\underbrace{\mathcal{R}^{d,\ell}\!\big(\hat{H}^{(\ell)}\big)}_{\text{upsample / transposed stride}},\qquad
\hat{H}^{(\ell-1)}=\underbrace{\mathcal{T}^{d,\ell}\!\big(\tilde{H}^{(\ell-1)};\boldsymbol{\theta}^{d,\ell}_T\big)}_{\text{conv/BN/activation}}.
\end{equation}
Finally, $\hat{\boldsymbol{\phi}}^i=\mathrm{reshape}^{-1}\!\big(\hat{H}^{(0)}\big)\in\mathbb{R}^{m}$. (Skip connections, if used, are part of $\mathcal{T}^{e,\ell}$/$\mathcal{T}^{d,\ell}$.) If using ConvTranspose layers, $\mathcal{R}^{d,\ell}$ and $\mathcal{T}^{d,\ell}$ can be implemented as a single deconvolutional operator.

\paragraph{Optimization Problem}
The optimization problem for CNN-AE is consistent with the general framework presented in  equation~\ref{eq:optimization_general}. The model parameters, which include all weights and  biases within the encoder and decoder networks  $\left\{\boldsymbol{\theta}_T^e, \boldsymbol{\theta}_R^e, \boldsymbol{\theta}_T^{d}, \boldsymbol{\theta}_R^{d}\right\}$, are learned by minimizing  the Mean-Squared Error (MSE) between the original and reconstructed fields. Unlike POD,  which admits a closed-form solution via linear algebra factorization (i.e., SVD), CNN-AE training requires an iterative  optimization procedure, typically carried out using stochastic gradient descent (SGD) or its variants.
    
\subsection{Conditional neural fields (CNF)}

\subsubsection{Neural field representation}
Neural fields (also called INRs) are continuous functions parameterized by neural networks that map spatial coordinates to physical quantities of interest. Given coordinates of $n_v$ points in $d$-dimensional space, $\mathbf{X} \in \mathbb{R}^{n_v\times d}$ a single-snapshot field is represented as,
\begin{equation}
    \boldsymbol{\phi}\ \approx\ f(\mathbf{X};\,\boldsymbol{\theta}),
\end{equation}
where $f(\cdot;\boldsymbol{\theta})$ is a neural network with parameters $\boldsymbol{\theta}$. Once trained, the continuity of $f$ enables evaluation at arbitrary $\mathbf{x}$, supporting interpolation and super-resolution.

\subsubsection{CNF for dimension reduction}
A naive extension to multiple snapshots would train one NF per snapshot, $f(\cdot;\boldsymbol{\theta}^i)$, which is computationally costly and ignores redundancy across time. In this work, we design a spatial dimension-reduction method that uses a {single} conditional neural field (CNF) as the decoder and {auto-decoding}~\cite{bojanowski2018optimizing,park2019deepsdf} to obtain latents. Specifically, a shared base network $f(\cdot;\boldsymbol{\theta})$ captures global structure, and a low-dimensional latent $\mathbf{z}^i\in\mathbb{R}^{r}$ modulates a subset of parameters via a linear/nonlinear projection:
\begin{equation}\label{eq:cnf_form}
    \boldsymbol{\phi}^i \ \approx\ f\!\big(\mathbf{X}^i;\,\boldsymbol{\theta}+\Delta\boldsymbol{\theta}(\mathbf{z}^i)\big),
\end{equation}
with $\mathbf{X}^i\in\mathbb{R}^{n_v^i\times d}$. We denote by $h$ the number of scalars actually modulated; typically $\Delta\boldsymbol{\theta}(\mathbf{z}^i)=\mathbf{M}\mathbf{z}^i$ with $\mathbf{M}\in\mathbb{R}^{h\times r}$, possibly organized per layer (block-diagonal, low-rank, or hypernetwork-generated). CNFs thus realize spatial dimension reduction by mapping each high-dimensional snapshot to a compact latent $\mathbf{z}^i$ while retaining an explicit, continuous decoder back to physical space. 

\paragraph{Encoding operation (auto-decoder)}
\begin{itemize}
\item \emph{Transformation step} ($\mathcal{T}^{e}$): 
Conceptually, fitting an NF to $(\mathbf{X}^i,\boldsymbol{\phi}^i)$ yields an implicit parameter vector $\boldsymbol{\gamma}^i$ in parameter space that best matches the snapshot:
\begin{equation}
\boldsymbol{\gamma}^i \;=\; \mathcal{T}^e\!\left(\mathbf{X}^i,\boldsymbol{\phi}^i ; \boldsymbol{\theta}_T^{e}=\emptyset\right)
\;=\; \arg \min_{\tilde{\boldsymbol{\gamma}}}\ \big\|\boldsymbol{\phi}^i-f\!\left(\mathbf{X}^i;\tilde{\boldsymbol{\gamma}}\right)\big\|_2^2.
\label{eq:cnf_encoder_transform}
\end{equation}
We do not solve \eqref{eq:cnf_encoder_transform} explicitly; it formalizes that the encoders intermediate representation $\boldsymbol{\gamma}^i$ lives in \emph{parameter space}.
\item \emph{Reduction step} ($\mathcal{R}^{e}$): 
We relate this implicit parameter vector to the latent via the linear map $\mathbf{M}$:
\begin{equation}
\mathbf{z}^i \;=\; \mathcal{R}^e\!\left(\boldsymbol{\gamma}^i; \boldsymbol{\theta}^e_R\right)
\;=\; \arg \min_{\mathbf{z}\in\mathbb{R}^{r}}\ \big\|\boldsymbol{\gamma}^i-\big(\boldsymbol{\theta}+\mathbf{M}\mathbf{z}\big)\big\|_2^2
\ \approx\ \mathbf{M}^{\dagger}\big(\boldsymbol{\gamma}^i-\boldsymbol{\theta}\big),
\label{eq:cnf_encoder_reduce}
\end{equation}
where $\mathbf{M}^{\dagger}$ denotes a (regularized) pseudoinverse. \emph{In practice we directly optimize $\mathbf{z}^i$ by minimizing reconstruction loss} (auto-decoding), which is numerically preferable to explicitly forming $\boldsymbol{\gamma}^i$:
\begin{equation}\label{eq:cnf_encode_opt}
\mathbf{z}^i \;=\; \arg\min_{\mathbf{z}\in\mathbb{R}^{r}}\
\mathcal{L}\!\left(\boldsymbol{\phi}^i,\ f\!\left(\mathbf{X}^i;\,\boldsymbol{\theta}+\mathbf{M}\mathbf{z}\right)\right).
\end{equation}
\end{itemize}
This realizes the encoder $\mathcal{E}$ as an \emph{optimization operator} $\mathcal{E}:\ (\mathbf{X}^i,\boldsymbol{\phi}^i)\mapsto \mathbf{z}^i$. (An amortized encoder $\mathcal{E}_\psi$ can be used instead; we adopt auto-decoding here.)

\paragraph{Decoding operation}
\begin{itemize}
\item \emph{Reverse reduction step} ($\mathcal{R}^{d}$):  
Given a low-dimensional latent vector $\mathbf{z}^i \in \mathbb{R}^{r}$, we expand it into higher-dimensional localized parameters corresponding to the snapshot-specific parameter shift $\boldsymbol{\theta} + \Delta \boldsymbol{\theta}^i$ via a linear transformation parameterized by $\mathbf{M} \in \mathbb{R}^{h\times r}$,
\begin{equation}
\hat{\boldsymbol{\gamma}}^i \;=\; \mathcal{R}^{d}\!\left(\mathbf{z}^i;\,\boldsymbol{\theta}_R^{d}\right)
\;=\; \boldsymbol{\theta}\;+\;\mathbf{M}\,\mathbf{z}^i \;\in\; \mathbb{R}^{h},
\label{eq:cnf_inverse_reduction}
\end{equation}
where $\boldsymbol{\theta}_R^{d}$ consists of the fixed base neural field parameters $\boldsymbol{\theta}$ and the learned linear mapping matrix $\mathbf{M}$. 
\item \emph{Reverse transformation step} ($\mathcal{T}^{d}$): 
The reverse transformation step for CNF is simply forward evaluation of the neural network, leading to the reconstructed field:
\begin{equation}
\hat{\boldsymbol{\phi}}^i \;=\; \mathcal{T}^{d}\!\left(\hat{\boldsymbol{\gamma}}^i;\,\boldsymbol{\theta}_T^{d}\right)
\;=\; f\!\left(\mathbf{X}^i;\,\hat{\boldsymbol{\gamma}}^i\right),
\end{equation}
where $\boldsymbol{\theta}_T^{d} = \emptyset$ contains no parameters, but solely the predefined computational operations within the base neural network.
\end{itemize}

\paragraph{Optimization problem}
The training objective jointly optimizes $(\boldsymbol{\theta},\mathbf{M})$ and the latents $\mathbf{Z}=\{\mathbf{z}^i\}_{i=1}^n$:
\begin{equation}
\label{eq:optimization_CNF_train}
\min_{\mathbf{Z},\,\boldsymbol{\theta},\,\mathbf{M}}\ \frac{1}{n}\sum_{i=1}^{n}
\mathcal{L}\!\left(\boldsymbol{\phi}^i,\ f\!\left(\mathbf{X}^i;\,\boldsymbol{\theta}+\mathbf{M}\mathbf{z}^i\right)\right)
\;+\;\lambda_z\,\frac{1}{n}\sum_{i=1}^{n}\|\mathbf{z}^i\|_2^2,
\end{equation}
where $\mathcal{L}$ is typically an $L^2$ or relative error on the field values, and $\lambda_z\!\ge\!0$ regularizes latents for stability/identifiability. 
At inference, we freeze the decoder and solve for the test latent:
\begin{equation}
\label{eq:optimization_CNF_infer}
\mathbf{z}_{\text{test}}^{\star} \;=\; \arg \min_{\mathbf{z}}\ \mathcal{L}\!\left(\boldsymbol{\phi}_{\text{test}},\ f\!\left(\mathbf{X}_{\text{test}};\,\boldsymbol{\theta}^{\star}+\mathbf{M}^{\star}\mathbf{z}\right)\right),
\end{equation}
and then reconstruct $\hat{\boldsymbol{\phi}}_{\text{test}}=f(\mathbf{X}_{\text{test}};\,\boldsymbol{\theta}^{\star}+\mathbf{M}^{\star}\mathbf{z}_{\text{test}}^{\star})$, where $(*)$ denotes the optimal values after training.

\paragraph{Remark}
Unlike a standard encoder–decoder, there is no explicit encoder $\mathcal{E}$ that maps high-dimensional fields $\boldsymbol{\Phi}$ to latents using \emph{auto-decoder} formulation.
Instead, the latents for the training snapshots, $\mathbf{Z}=\big[\mathbf{z}^1,\ldots,\mathbf{z}^n\big]$,  are introduced as free learnable variables and are optimized jointly with the shared base-network parameters $\boldsymbol{\theta}$ and the conditioning-module parameters. Thus, ``encoding'' is realized implicitly by optimization rather than by a separate operator. At inference, given an unseen snapshot, its latent $\mathbf{z}_{\text{test}}$ is obtained by solving a small optimization problem with base network held fixed.

\subsubsection{Conditioning mechanisms}

Conditioning specifies how auxiliary context modulates the decoder so that the predicted field depends on both spatial coordinates and context. We write the generic conditioned mapping as
\begin{equation}
    \hat{\boldsymbol{\phi}} = {f}(\mathbf{X},\mathbf{C};\theta)
    \label{eqn:conditioning}
\end{equation}
where $\mathbf{X}$ denotes spatial coordinates and $\mathbf{C}$ carries conditioning information. 
In our CNF, the conditioning variable is the latent $\mathbf{z}$; conditioning is realized either by \emph{parameter modulation} $\Delta\boldsymbol{\theta}(\mathbf{z})$ of the base network or via a \emph{last-layer coupling}  (DeepONet-style inner product) without modifying internal weights.

A simple historical baseline of conditioning is \emph{concatenation}, which appends $\mathbf{C}$ to the input or intermediate features. While easy to implement and inexpensive, concatenation typically induces only weak interactions between $\mathbf{X}$ and $\mathbf{C}$ and often underperforms when strong, structured coupling is required.

\begin{figure}[htp!]
    \centering
    \includegraphics[width=1\linewidth]{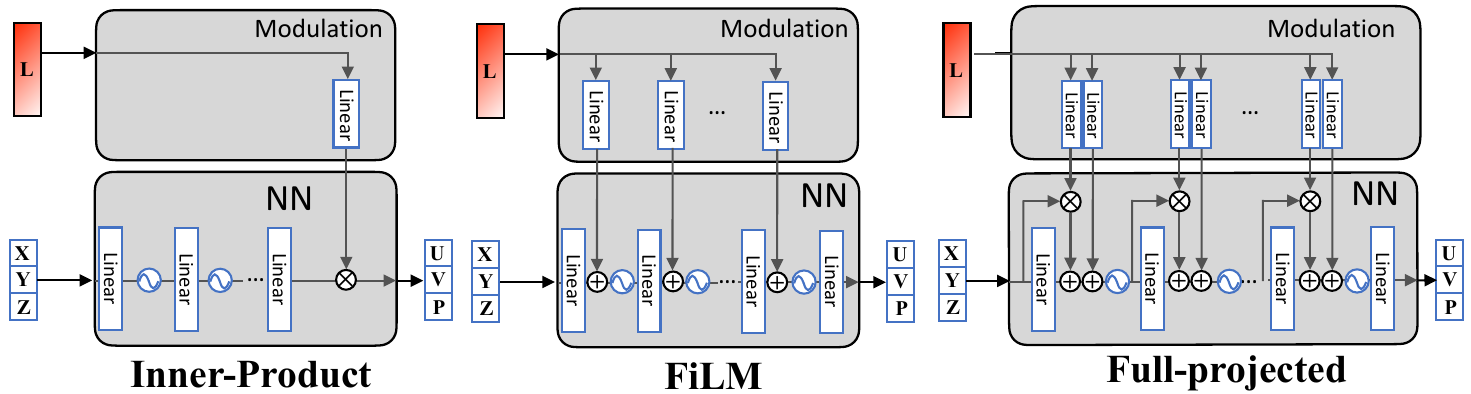}
    \caption{Diagram of different conditioning mechanism}
    \label{fig:conditioning}
\end{figure}

To obtain stronger inductive bias and controllable capacity, we adopt the three mechanisms illustrated in \Cref{fig:conditioning}. Below we give their layer-wise forms and brief context. 
Let layer $\ell$ have pre-activation $u^{(\ell)}=W^{(\ell)}h^{(\ell)}+b^{(\ell)}$, activation $h^{(\ell+1)}=\rho\!\big(u^{(\ell)}\big)$, widths $d_\ell,d_{\ell+1}$, and denote $M_B^{(\ell)}\!\in\!\mathbb{R}^{d_{\ell+1}\times r}$ and $M_W^{(\ell)}\!\in\!\mathbb{R}^{(d_{\ell+1}\times d_{\ell})\times r}$ as the latent-to-bias projection and the latent-to-weights projection respectively.

\begin{itemize}[leftmargin=1.2em]
    \item {CNF–FiLM (feature-wise linear modulation).} 
    FiLM-style activation modulation is widely used in conditional representation learning because it injects sample-dependent information with minimal memory/latency overhead and good training stability. In NFs, it preserves the base operator while allowing snapshot-specific shifts at each layer.
    Specifically, we keep weights fixed and modulate activations using latent-dependent shifts (and optionally gains). Our default bias-only variant reads
    \begin{equation}
        \label{eqn:film_layer_wise}
        h^{(\ell+1)} 
        \;=\; \rho\!\Big(W^{(\ell)}h^{(\ell)} + b^{(\ell)} + \underbrace{M_B^{(\ell)}\mathbf{z}}_{\Delta b^{(\ell)}(\mathbf{z})}\Big).
    \end{equation}
    Complexity is $O(d_{\ell+1}r)$ parameters per layer and negligible runtime overhead. We zero-initialize $M_B^{(\ell)}$ so that $\mathbf{z}=\mathbf{0}$ recovers the base network.
    \item {CNF–FP (full-projected weight{+}bias modulation).}
    We generalize FiLM by permitting latent-driven weight updates. It is analogous to hypernetwork-style adaptation used to increase expressivity while controlling parameter growth. The added flexibility often improves fit but can increase overfitting and computation overhead. We allow latent-driven updates of both weights and biases: 
    \begin{equation}
        \label{eqn:fp_layer_wise}
        \begin{aligned}
            u^{(\ell)}(\mathbf{z}) 
            &= \big(W^{(\ell)} + \Delta W^{(\ell)}(\mathbf{z})\big)\,h^{(\ell)}
             + \big(b^{(\ell)} + \Delta b^{(\ell)}(\mathbf{z})\big),\\
            \Delta b^{(\ell)}(\mathbf{z}) &= M_B^{(\ell)}\mathbf{z}, \qquad
            \Delta W^{(\ell)}(\mathbf{z}) 
            = M_W^{(\ell)}\mathbf{z}.
        \end{aligned}
    \end{equation}
     We zero-initialize $M_B^{(\ell)}$ and $M_W^{(\ell)}$  so that $\mathbf{z}=\mathbf{0}$ recovers the base network.

    \item {CNF–Inner (DeepONet-style last-layer coupling).}
    Inner-product coupling is standard in branch–trunk operator learning: a trunk encodes coordinates, a branch encodes context, and a final inner product yields the output. It is memory/latency efficient and naturally rank-controlled by $r$, but lacks internal parameter adaptation. 
    Internal weights are not modulated. A trunk produces $\psi(\mathbf{x})\in\mathbb{R}^{r}$ and a branch maps the latent $b(\mathbf{z})\in\mathbb{R}^{r}$ (often $b(\mathbf{z})=\mathbf{z}$ or $b(\mathbf{z})=\mathbf{B}\mathbf{z}$); the output is
    \begin{equation}
        f(\mathbf{x};\mathbf{z}) \;=\; \psi(\mathbf{x})^{\top} b(\mathbf{z})
        \quad \big(\text{or } f(\mathbf{x};\mathbf{z})=\Psi(\mathbf{x})\,b(\mathbf{z}) \text{ for multi-channel outputs}\big).
    \end{equation}
\end{itemize}

CNF–FiLM and CNF–FP implement the decoder's reverse-reduction 
$\mathcal{R}^{d}:\mathbf{z}\mapsto \boldsymbol{\theta}+\Delta\boldsymbol{\theta}(\mathbf{z})$ 
in parameter space, followed by forward evaluation $\mathcal{T}^{d}$, whereas CNF–Inner realizes conditioning entirely inside $\mathcal{T}^{d}$ with a minimal $\mathcal{R}^{d}$. 
These mechanisms span a practical capacity–cost trade space, and we benchmark all three in Section~\ref{sec:Result}.

\subsubsection{Domain-decomposition for CNFs}

High-fidelity turbulent snapshots routinely contain $10^6$–$10^8$ spatial degrees of freedom. With $\mathcal{O}(10^2$-$10^3)$ snapshots available for training, a single global CNF must explain all multi-scale variability with one latent per snapshot and therefore tends to overfit and generalize poorly. To improve the bias-variance tradeoff, we introduce a \emph{domain-decomposed CNF} that conditions locally while sharing a global base network. 

Let the domain be partitioned into $P$ subdomains (patches)
$\{\Omega_p\}_{p=1}^{P}$ with $\bigcup_{p=1}^{P}\Omega_p=\Omega$ (non–overlapping in our default,
though the formulation also supports overlaps). For snapshot $i$ we associate a
\emph{patch latent} $\mathbf{z}^i_p\in\mathbb{R}^{r}$ with each $\Omega_p$.
All patches share the same decoder parameters $\boldsymbol{\theta}$ and conditioning
projection $\mathbf{M}\in\mathbb{R}^{h\times r}$ learned from data. We also define a local
coordinate normalization $\widetilde{\mathbf{x}}=\mathcal{N}_p(\mathbf{x})\in[-1,1]^d$ for
$\mathbf{x}\in\Omega_p$ to improve conditioning of the SIREN/MLP. 
Within patch $\Omega_p$ the decoder evaluates
\begin{equation}
\label{eq:dd_local_decode}
\hat{\phi}^i_p(\mathbf{x})
=
f\big(\widetilde{\mathbf{x}};\,\boldsymbol{\theta}+\Delta\boldsymbol{\theta}(\mathbf{z}^i_p)\big),
\qquad
\Delta\boldsymbol{\theta}(\mathbf{z}^i_p)=\mathbf{M}\mathbf{z}^i_p,
\quad \mathbf{x}\in\Omega_p,
\end{equation}
i.e., the localized reverse–reduction $\mathcal{R}^{d}:\mathbf{z}^i_p\to
\boldsymbol{\theta}+\mathbf{M}\mathbf{z}^i_p$ followed by the forward evaluation
$\mathcal{T}^{d}$. For a non–overlapping tiling we
assemble the global prediction by restriction,
$\hat{\phi}^i(\mathbf{x})=\hat{\phi}^i_p(\mathbf{x})$ if $\mathbf{x}\in\Omega_p$.
For overlapping tiles, weighted summation is needed (not pursued in this study). 

Let $\mathcal{S}^i_p\subset\Omega_p$ be the training samples for patch $p$ in snapshot $i$.
We minimize the sum of patchwise reconstruction errors, with optional interface and smoothness
regularization:
\begin{equation}
\label{eq:dd_loss}
\begin{aligned}
\min_{\boldsymbol{\theta},\,\mathbf{M},\,\{\mathbf{z}^I_p\}}\;
&\sum_{i}\sum_{p}\;
\mathbb{E}_{\mathbf{x}\in\mathcal{S}^I_p}
\Big\|\phi^I(\mathbf{x})-\hat{\phi}^I_p(\mathbf{x})\Big\|_2^2
\;+\;
\lambda_{\mathrm{int}}\sum_{i}\sum_{(p,q)\in\mathcal{E}}\;
\mathbb{E}_{\mathbf{x}\in\mathcal{S}^I_{pq}}
\Big\|\hat{\phi}^I_p(\mathbf{x})-\hat{\phi}^I_q(\mathbf{x})\Big\|_2^2
\\
&\;+\;
\lambda_{\mathrm{spa}}\sum_{i}\sum_{(p,q)\in\mathcal{E}}
\big\|\mathbf{z}^I_p-\mathbf{z}^I_q\big\|_2^2
\;+\;
\lambda_{\mathrm{tem}}\sum_{p,i}\big\|\mathbf{z}^{i+1}_p-\mathbf{z}^I_p\big\|_2^2,
\end{aligned}
\end{equation}
where $\mathcal{E}$ indexes neighboring patches, $\mathcal{S}^I_{pq}\subset\Omega_p\cap\Omega_q$
samples the interface (used only if seam suppression is desired), and
$\lambda_{\mathrm{int}},\lambda_{\mathrm{spa}},\lambda_{\mathrm{tem}}\ge 0$ control the strength of the regularizers.

The parameter count of the shared decoder $(\boldsymbol{\theta},\mathbf{M})$ is independent of $P$; only the per–snapshot latents scale as $O(Pr)$. Smaller patches reduce local complexity and improve extrapolation (especially near walls) at the cost of more latents and patch evaluations; In practice, we choose uniform tiling of equal-sized patches and train with balanced mini-batches across patches. While anisotropic tiling is compatible with our formulation and may further improve accuracy for inhomogeneous flows, but we leave such adaptivity to future work.

\section{Numerical Results}
\label{sec:Result}

\subsection{Experimental setup}
\label{subsec:Experiment_setting}

We evaluate spatial dimension reduction and reconstruction across time by training on a subset of snapshots and assessing accuracy on \emph{unseen} snapshots drawn from two disjoint regimes: \emph{interpolation} (in-range) and \emph{extrapolation} (out-of-range). Let $\boldsymbol{q}(\mathbf{x},t)$ denote the spatiotemporal field with spatial coordinate $\mathbf{x}\in\Omega$ and time $t\in[0,T{+}T')$.

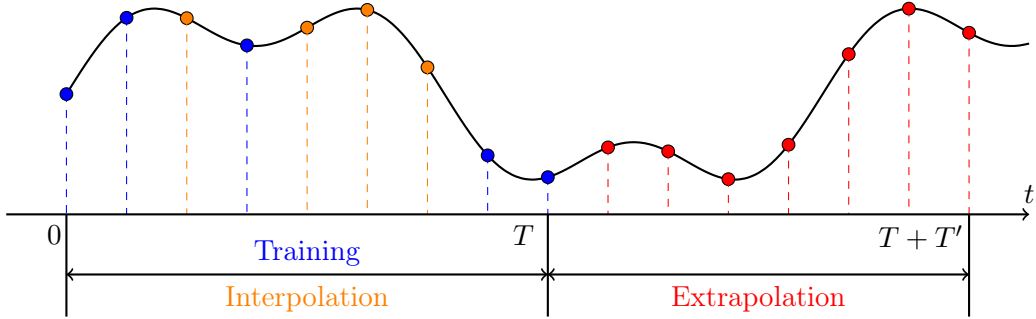
\begin{figure}[h!]
    \centering
    \begin{tikzpicture}[font=\small, scale=1.6]
    
    % --- 1. Draw a "random" continuous signal ---
    % For demonstration, use a sinusoidal combination:
    \draw[thick, black, domain=0:8, samples=100, smooth] 
      plot(\x, {0.7*sin(\x r) + 0.3*sin(3 * \x r) + 1.0});
    
    % --- 2. Mark training points ---
    \foreach \x in {0, 0.5,1.5,3.5,4.0} {
      \pgfmathparse{0.7*sin(\x r) + 0.3*sin(3*\x r) + 1.0}
      \let\y=\pgfmathresult
      \draw[fill=blue] (\x, \y) circle (1.5pt);
      \node[blue, above] at (\x, \y) {} ;
      
      % Add a dashed line starting from the sample point
      \draw[dashed,blue] (\x, \y) -- (\x, 0);
    }
    
    % --- 3. Mark interpolation points (between training points) ---
    \foreach \x in {1.0, 2, 2.5,3} {
      \pgfmathparse{0.7*sin(\x r) + 0.3*sin(3*\x r) + 1.0}
      \let\y=\pgfmathresult
      \draw[fill=orange] (\x, \y) circle (1.5pt);
      \node[orange, above] at (\x, \y) {};
        % Add a dashed line starting from the sample point
      \draw[dashed,orange] (\x, \y) -- (\x, 0);
    }
    
    % % --- 4. Mark extrapolation points (outside training interval) ---
    \foreach \x in {4.5,5,5.5, 6,6.5,7,7.5} {
      \pgfmathparse{0.7*sin(\x r) + 0.3*sin(3*\x r) + 1.0}
      \let\y=\pgfmathresult
      \draw[fill=red] (\x, \y) circle (1.5pt);
      \node[red, above] at (\x, \y) {} ;
        \draw[dashed,red] (\x, \y) -- (\x, 0);
    }
    
    % --- Add base x-axis ---
    
    \draw[->, thick] (-0.5, 0) -- (8, 0) node[anchor=north, above] {$t$};
    
    % --- Add label below x-axis ---
    \draw[<->, thick] (0, -0.5) -- (4, -0.5) 
        node[midway, above,blue,yshift=-1pt] {Training}
        node[midway, below,orange] {Interpolation};
    \draw[<->, thick] (4, -0.5) -- (7.5, -0.5) node[midway, below, red] {Extrapolation};
    
    \draw[thick] (0,0) -- (0,-0.85){};
    \draw[thick] (4,0) -- (4,-0.85){};
    \draw[thick] (7.5,0) -- (7.5,-0.85){};

    % --- Add label below x-axis ---
    \node[below] at (3.8, 0) {$T$};
    \node[below] at (7.1, 0) {$T+T'$};
    \node[below] at (-0.1,0) {$0$};
    \end{tikzpicture}

    \caption{Demo diagram of dataset splitting strategy}
    \label{fig:enter-label}
\end{figure}

The training dataset, $\Dcal_{\text {train }}$, is defined as:
\begin{equation}
    \Dcal_{\text {train }}=\left\{\qvb(\mathbf{x},t) \mid t \in \Tcal_{\text {train }}, \mathbf{x} \in \Omega \right\},
\end{equation}
where $\Tcal_{\text {train }} \subset[0, T)$ is a randomly sampled subset of the time domain $[0, T)$. This dataset is used to construct the dimension reduction models.

To evaluate in-range generalization (interpolation within the training horizon), we define the interpolative test set as:
\begin{equation}
    \mathcal{D}_{\text {interp }}=\left\{\mathbf{q}(\mathbf{x}, t) \mid t \in \Tcal_{\text {interp }}, \mathbf{x} \in \Omega\right\},
\end{equation}
where $\Tcal_{\text{interp}}$ is a set of time indices drawn from $[0,T)$ with
$\Tcal_{\text{train}}\cap\Tcal_{\text{interp}}=\varnothing$. Thus, every interpolation snapshot is \emph{unseen} yet lies within the training time span $[0,T)$ (in-distribution), i.e., at disjoint indices but not beyond the training horizon.

To assess out-of-range generalization beyond the training horizon, we define the extrapolative dataset as:
\begin{equation}
    \Dcal_{\text {extrap }}=\left\{\qvb(\mathbf{x},t) \mid t \in\left[T, T+T^{\prime}\right), \mathbf{x} \in \Omega\right\}.
\end{equation}
Unless otherwise noted, all quantitative claims of ``extrapolation'' refer strictly to $\Dcal_{\text{extrap}}$.

Most ML-based dimensionality-reduction studies for turbulence evaluate on \emph{interpolative} test sets, held-out samples whose indices remain within the training range, rather than on truly out-of-range data. In such settings, test snapshots are \emph{in-distribution}, which partially explains why many methods report strong performance. To make this distinction explicit and fair, we evaluate and report both protocols side by side, using identical preprocessing, metrics, and evaluation grids across splits; the same $\big(\Tcal_{\text{train}},\Tcal_{\text{interp}},\Tcal_{\text{extrap}}\big)$ indices are fixed and shared across all baselines for comparability.

We consider three datasets: (i) velocity fields on planes perpendicular to the streamwise direction from turbulent channel flows (DNS), (ii) wall pressure fluctuations over a turbulent flat boundary layer, and (iii) inlet streamwise velocity for turbulent channel flows from wall modeled LES (WMLES). Each dataset includes uniformly sampled time snapshots over $[0,T{+}T')$ and fields stored on their native evaluation grids. Details of data generation (governing equations, Reynolds numbers, numerical schemes, discretizations, sampling cadence, and boundary conditions) are provided in \ref{app:data}.

\subsection{Benchmark study against linear and DL baselines}
\label{sec:benchmark_study}
We first compare CNFs with different conditioning mechanisms (FiLM, FP, and inner product) to widely used dimensionality-reduction baselines under identical splits and metrics on the WMLES--Inlet dataset. Table~\ref{tab:Inlet_LES_error} reports relative $L^2$ errors across latent sizes $r\in\{8,16,32,64,128,256\}$ and splits (training, interpolation, strict extrapolation). 

\begin{table}[t!]\centering
\scriptsize
\begin{tabular}{l l c c c c c}\toprule
\textbf{Latent size} & \textbf{Split} & \textbf{POD} & \textbf{ConvAE} & \textbf{CNF-FP} & \textbf{CNF-FiLM} & \textbf{CNF-Inner} \\ \midrule

\multirow{3}{*}{8} 
& Training        & 4.78\%                & 3.82\%                & \textbf{1.17\%}          & 2.48\%        & 5.66\%      \\
& Interpolation  & 4.83\%                & 4.04\%                & \textbf{2.27\%}          & 3.14\%        &  5.58\%     \\
& Extrapolation  & \textbf{5.74\%}                & 6.14\%                & 6.60\%          & 6.22\%        &  6.00\%     \\ \midrule

\multirow{3}{*}{16} 
& Training        & 4.21\%                & 3.28\%                & \textbf{0.81\%}          & 2.26\%        &  5.65\%    \\
& Interpolation  & 4.32\%                & 3.50\%                & \textbf{1.06\%}          & 2.60\%        &  5.66\%   \\
& Extrapolation  & \textbf{5.49\%}                & 5.87\%                & 6.01\%          & 5.80\%        &  6.02\%    \\ \midrule

\multirow{3}{*}{32} 
& Training        & 3.46\%                & 2.86\%                & \textbf{0.46\%}          & 2.10\%        &  5.65\%    \\
& Interpolation  & 3.64\%                & 3.10\%                & \textbf{0.73\%}          & 2.37\%        &  5.66\%    \\
& Extrapolation  & \textbf{5.01\%}                & 5.43\%                & 5.21\%          & 5.05\%        &  6.02\%   \\ \midrule

\multirow{3}{*}{64} 
& Training        & 2.54\%                & 2.19\%                & \textbf{0.39\%}          & 1.73\%        &  5.65\%   \\
& Interpolation  & 2.81\%                & 2.45\%                & \textbf{0.54\%}          & 2.02\%        &  5.66\%   \\
& Extrapolation  & 4.48\%                & 4.98\%                & 4.37\%          & \textbf{4.12\%}        &  6.02\%     \\ \midrule

\multirow{3}{*}{128} 
& Training        & 1.52\%                & 1.76\%                & \textbf{0.21\%}          & 1.37\%        &  5.65\%   \\
& Interpolation  & 1.85\%                & 1.99\%                & \textbf{0.36\%}          & 1.64\%        &  5.66\%  \\
& Extrapolation  & 3.80\%                & 4.80\%                & 3.61\%          & \textbf{3.10\%}        &  6.02\%  \\ \midrule

\multirow{3}{*}{256} 
& Training        & 0.55\%                & 1.39\%                & \textbf{0.10\%}          & 1.21\%        &  5.65\%    \\
& Interpolation  & 0.84\%                & 1.65\%                & \textbf{0.28\%}          & 1.36\%        &  5.66\%   \\
& Extrapolation  & 3.09\%                & 4.82\%                & 2.82\%          & \textbf{2.04\%}        &  6.01\%     \\ \bottomrule
\end{tabular}
\caption{Relative $L^2$ errors (\%) on the WMLES-Inlet dataset across latent sizes and evaluation splits.}
\label{tab:Inlet_LES_error}
\end{table}

Focusing on \emph{fitting capability} (training), errors decrease with latent size for all methods, and CNF--FP is the most accurate at every $r$ (e.g., $0.46\%$ at $r{=}32$, $0.21\%$ at $r{=}128$, $0.10\%$ at $r{=}256$), followed by CNF--FiLM ($2.10\%$, $1.37\%$, $1.21\%$) and ConvAE ($2.86\%$, $1.76\%$, $1.39\%$); POD improves to $0.55\%$ at $r{=}256$ but remains above CNF--FP, and CNF--Inner stays near $5.65\%$, indicating persistent underfitting. On \emph{in-distribution testing} (interpolation), the ranking persists: CNF--FP attains the lowest errors throughout, CNF--FiLM is next, then ConvAE; POD improves with capacity, while CNF--Inner remains $\approx 5.66\%$. For \emph{out-of-range testing}, the picture changes: once $r \ge 64$, CNF--FiLM provides the best accuracy among learnable decoders, with errors $4.12\%$ ($r{=}64$), $3.10\%$ ($r{=}128$), and $2.04\%$ ($r{=}256$), outperforming CNF--FP ($4.37\%$, $3.61\%$, $2.82\%$), ConvAE ($4.98\%$, $4.80\%$, $4.82\%$, which plateaus), and POD at the same $r$; at very small latent sizes ($r \le 32$) the linear POD baseline is competitive in $L^2$, consistent with its bias toward energetic low–wavenumber content. Generalization gaps at $r{=}128$ further quantify the trade-off: extrapolation minus training increases by $+1.73$ for CNF--FiLM ($1.37 \to 3.10\%$) versus $+3.40$ for CNF--FP ($0.21 \to 3.61\%$), with POD and ConvAE in between ($+2.28$ and $+3.04$); thus, while CNF--FP offers the strongest data fitting and in-range reconstruction, activation-only modulation (CNF--FiLM) yields more stable out-of-range generalization once moderate capacity is available. Figure~\ref{fig:snapshot_LES_Inlet_combined} in Appendix complements these findings with side-by-side reconstructions and absolute-error maps at $r{=}128$. Consistent with the quantitative results, CNF--FiLM presents the best out-of-range performance with reduced small-scale errors, CNF--FP is visually sharp in-range but degrades more beyond the horizon, POD appears smoother, and ConvAE sits between POD and CNFs.

\begin{figure}[t!]
    \centering
    \includegraphics[width=1\linewidth]{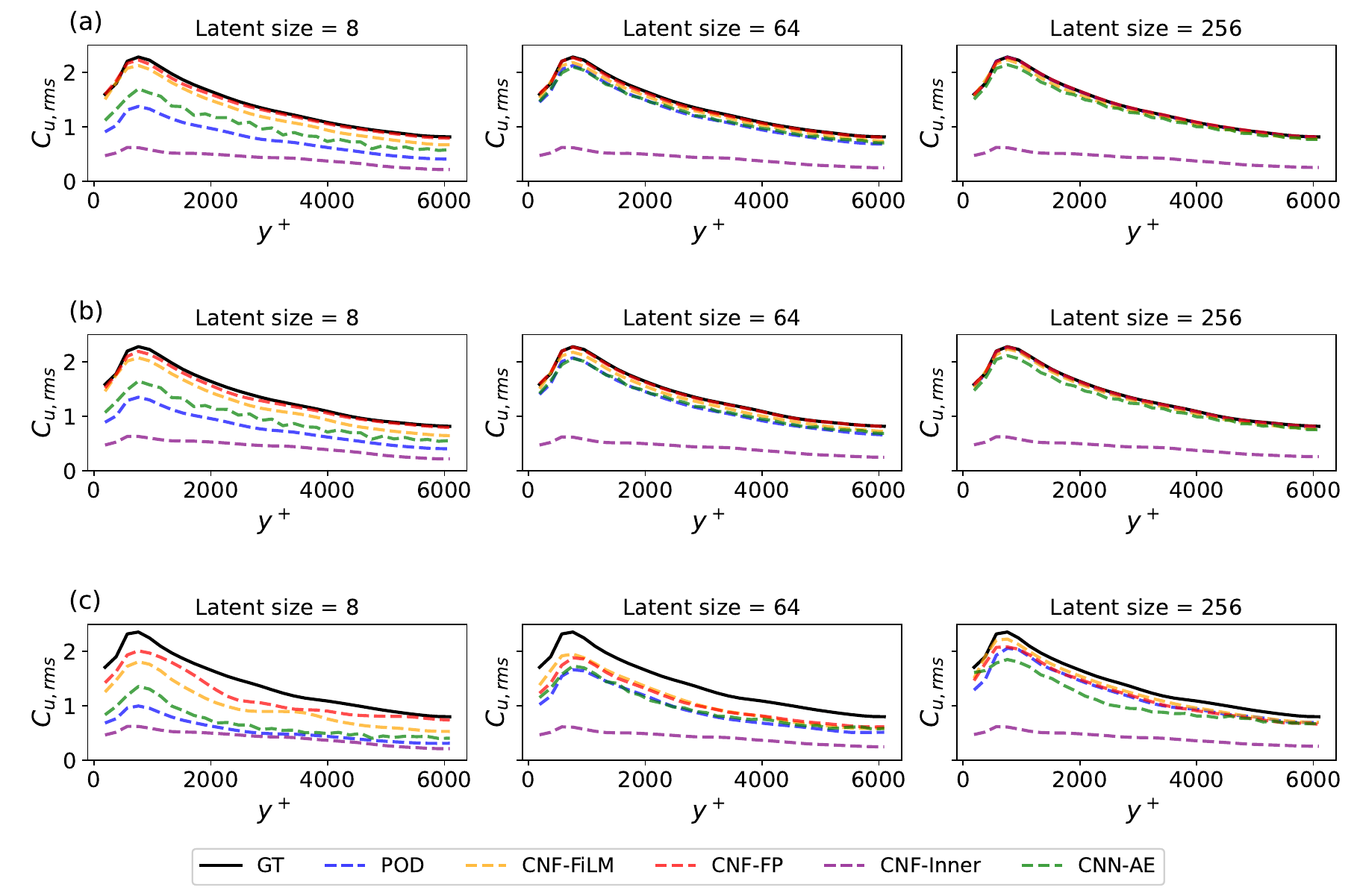}
    \caption{Wall–normal profiles of the normalized streamwise–velocity fluctuation RMS,
$C_{u,\mathrm{rms}}(y^+)$, on the WMLES--Inlet dataset. Columns correspond to latent sizes $r \in\{8,64,256\}$; rows shows evaluation splits: (a) training, (b) in-range testing, (c) out-of-range testing. }
    \label{fig:LES_inlet_stats}
\end{figure}

While the relative $L^2$ error comparisons quantify pointwise reconstruction accuracy, they are agnostic to flow physics and can obscure scale-dependent errors. We therefore investigate the turbulence statistics of the reconstructed fields. Figure~\ref{fig:LES_inlet_stats} presents wall-normal profile of root mean square (rms) of streamwise–velocity fluctuations $C_{u,\mathrm{rms}}(y^+)$ for training, interpolation, and extrapolation at representative latent sizes, providing a physics-grounded view of how each model recovers the near-wall peak and outer-layer decay.
Increasing the latent dimension improves agreement with the ground truth in all regimes: the near-wall peak and the outer-layer decay are progressively recovered as $r$ grows from $8$ to $256$. In the \emph{out-of-range testing} panel (Fig.~\ref{fig:LES_inlet_stats}c), POD exhibits the largest deficit across $y^+$, particularly around the peak region, despite its relatively low $L^2$ error at small $r$ (Table~\ref{tab:Inlet_LES_error}); this reflects POD's bias toward energetic low-wavenumber content that suppresses small-scale intensity. By contrast, CNF decoders move closer to the ground truth with capacity: at $r = 64$ and $256$, CNF--FiLM tracks both the magnitude and the position of the $C_{u,\mathrm{rms}}$ peak most closely under extrapolation, while CNF--FP, which is the best fitter on the training and in-distribution testing, retains a small but visible underprediction in the outer region. ConvAE improves with $r$ yet consistently underperforms than CNFs (FiLM and FP), and CNF--Inner changes little with capacity, consistent with its flat $L^2$ performance. Taken together, these diagnostics confirm that the apparent $L^2$ advantage of linear POD at small $r$ does not translate to physically faithful fluctuation levels, whereas CNF–FiLM offers the most robust recovery of wall-normal turbulence intensity beyond the training horizon.

\subsection{CNF with domain decomposition}

Building on the WMLES--Inlet benchmark where CNFs outperformed or matched POD and ConvAE, we now assess whether a \emph{single global latent} remains sufficient for more demanding data. Two cases increase difficulty substantially relative to WMLES: (i) DNS-resolution inlet slices of channel flow, which contain richer small-scale content, and (ii) instantaneous wall-pressure fluctuations over a flat boundary layer, whose signal is intermittent and broadband. In both settings a global CNF underresolves fine structures and degrades out of range, whereas introducing \emph{domain decomposition}, one latent per spatial patch with a shared decoder, recovers sharpness and improves generalization. The effectiveness of the proposed domain-decomposed CNFs is systematically assessed through visual comparisons of snapshot reconstructions, quantitative error analysis, and evaluations of turbulence statistics.

\subsubsection{Inflow turbulence of DNS channel flows}

\begin{figure}[htp!]
    \centering
    \includegraphics[width=0.8\linewidth]{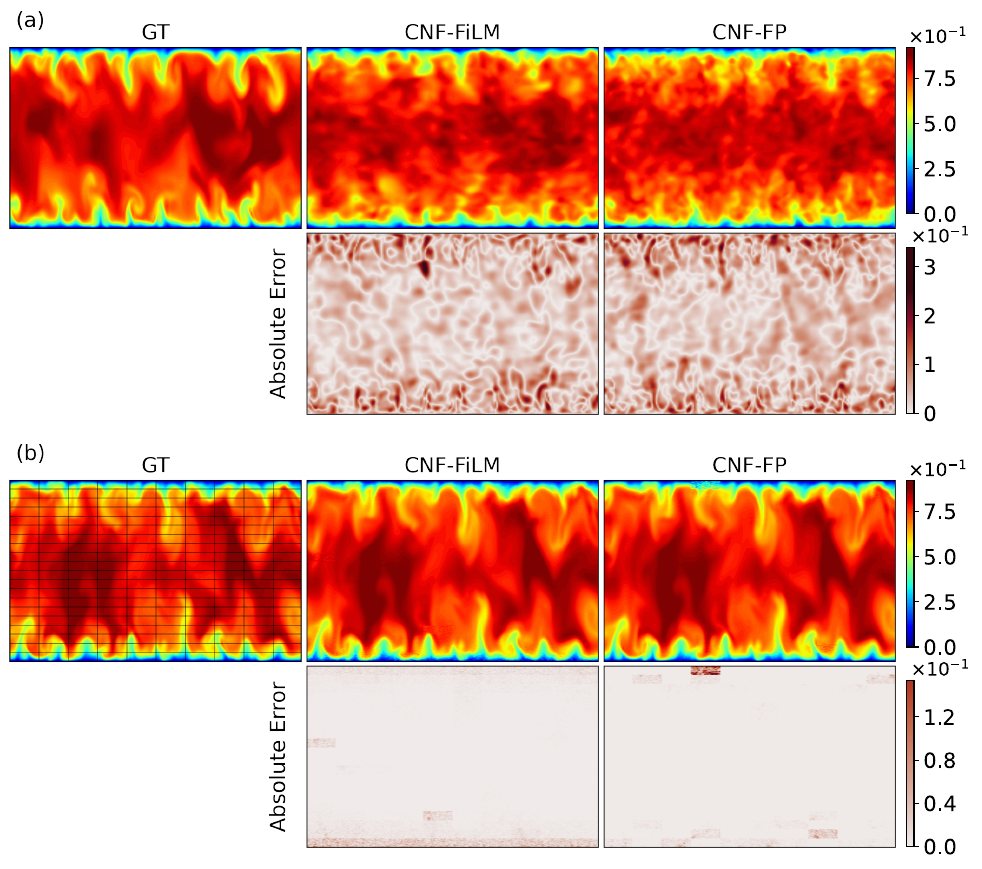}
    \caption{DNS-inlet, out-of-range testing at $r = 128$. Top: reconstructions; bottom: absolute error. (a) Global CNFs (no decomposition): blurred streaks, spurious high–wavenumber textures, larger structured errors. (b) Domain-decomposed CNFs: streak spacing and amplitude recovered; artifacts suppressed.}
    \label{fig:snapshot_extrap_DNSInlet}
\end{figure}

When the grid is refined to DNS resolution, the instantaneous velocity fields carry a much broader spectral bandwidth and sharper gradients, and near-wall viscous streaks coexist with outer-layer energy–containing motions, so a single global latent struggles to represent all scales. The limitation is most evident under out-of-range testing: Fig.~\ref{fig:snapshot_extrap_DNSInlet} presents out-of-range reconstructions at $r{=}128$ (global CNFs vs.\ domain-decomposed CNFs with FiLM/FP).

Without decomposition, both CNF--FiLM and CNF--FP blur core-region streaks, attenuate small-scale contrast, underresolve near-wall modulation, and introduce speckle-like high-wavenumber artifacts absent in the DNS. With decomposition, streak spacing and amplitude are largely restored and spurious fine-scale textures are suppressed. 

Table~\ref{tab:Inlet_DNS_error} quantifies the reconstruction errors for both training and out-of-range testing scenarios. Across latent sizes $r\in\{32,64,128\}$, domain decomposition significantly reduces strict-extrapolation error for both conditioning mechanisms. Notably, the reconstruction error of decomposed CNF-FiLM model on out-of-range testing samples drops from 6.64\% to 1.11\% at $r{=}128$, with similarly large reductions at $r{=}32$ and $64$, indicating that local conditioning primarily improves out-of-range robustness.
\begin{table}[!thb]\centering
\scriptsize
\begin{tabular}{l l c c c c}
\toprule
& & \multicolumn{2}{c}{\textbf{With Decomposition}} & \multicolumn{2}{c}{\textbf{Without Decomposition}} \\
\cmidrule(lr){3-4}\cmidrule(lr){5-6}
\textbf{Latent size} & \textbf{Dataset} & \textbf{CNF-FP} & \textbf{CNF-FiLM} & \textbf{CNF-FP} & \textbf{CNF-FiLM} \\
\midrule
\multirow{2}{*}{32}
& Training &  \textbf{0.35}\% & 0.73\% & 1.25\% & 4.03\% \\
& Testing extrap & \textbf{0.79}\% & 0.96\% & 9.66\% & 9.33\% \\
\midrule
\multirow{2}{*}{64}
    & Training & \textbf{0.13}\% & 0.55\% & 0.88\% & 3.59\% \\
& Testing extrap & 3.03\% & \textbf{0.76}\% & 8.67\% & 8.28\% \\
\midrule
\multirow{2}{*}{128}
& Training & \textbf{0.06}\% & 0.48\% & 0.27\% & 2.93\% \\
& Testing extrap & 1.37\% & \textbf{1.11}\% & 7.28\% & 6.64\% \\
\bottomrule
\end{tabular}
\caption{Relative $L^2$ errors (\%) on the DNS-Inlet dataset, comparing CNFs \emph{with} and \emph{without} domain decompositions.}
\label{tab:Inlet_DNS_error}
\end{table}

\begin{figure}[t!]
    \centering
    \includegraphics[width=1\linewidth]{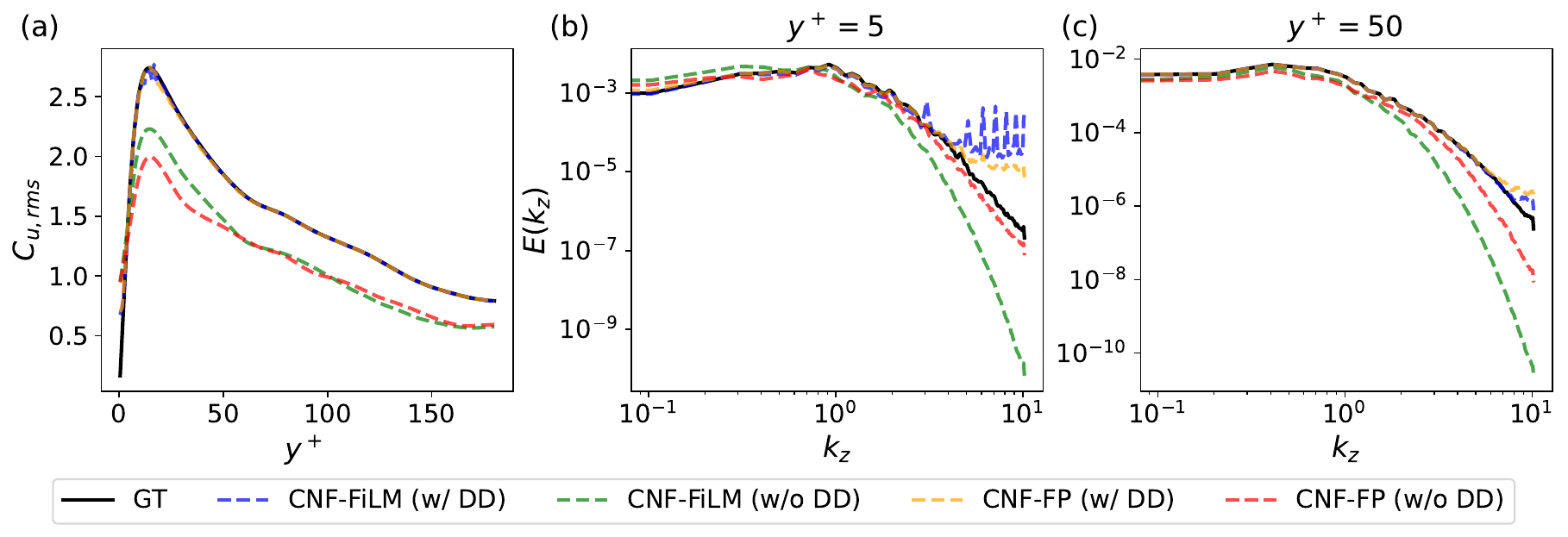}
    \caption{Turbulence statistics of reconstructed out-of-range testing samples at $r = 128$. (a) Wall-normal profile of streamwise–velocity fluctuations $C_{u,\mathrm{rms}}(y^+)$. (b,c) Spanwise energy spectra $E(k_z)$ at near-wall ($y^+=5$) and outer-layer ($y^+=50$) locations.}
    \label{fig:DNS_Inlet_stats_extrap}
\end{figure}

Further statistical comparisons are provided in Figure~\ref{fig:DNS_Inlet_stats_extrap}. Panel (a) shows the wall-normal distribution of streamwise–velocity fluctuations $C_{u,\mathrm{rms}}(y^+)$, where domain-decomposed CNFs (w/ DD) reproduce both the near-wall peak and the outer-layer decay with high fidelity, while global CNF models (w/o DD) systematically underpredict fluctuation intensity, consistent with the instantaneous snapshot visualizations. Figures~\ref{fig:DNS_Inlet_stats_extrap}b and \ref{fig:DNS_Inlet_stats_extrap}c present the spanwise energy spectra $E(k_z)$ at $y^+ = 5$ (near-wall) and $y^+ = 50$ (outer layer), respectively. At both locations, models with domain decomposition align markedly better with the DNS spectrum, sustaining the inertial-range slope and delaying spectral roll-off to higher $k_z$. By contrast, global CNFs exhibit a clear high-$k_z$ energy deficit, with CNF--FiLM (w/o DD, green) worst and CNF--FP (w/o DD, red) somewhat better yet still biased low. However, at the very near-wall region ($y^+ = 5$), the domain decomposition introduces a mild overshoot/oscillation at high $k_z$ (most visible for CNF--FiLM w/ DD), indicative of slight over-amplification of small-scale energy at patch interfaces. Overall, domain-decomposed CNF substantially corrects the spectral bias of global CNFs while introducing only small, localized ripples that are negligible at $y^+ = 50$ and can be mitigated with overlap-and-blend or interface regularization.

%The wall-normal coordinate $y^+$ is defined as $y^+ = y u_{\tau} / \nu$, where $y$ is the distance from the wall, $u_{\tau}$ is the friction velocity, and $\nu$ is the kinematic viscosity. This nondimensional parameter characterizes the wall distance in viscous units. The decomposed CNF-FiLM model consistently provides the closest match to the ground truth across the spectrum. A common characteristic observed across all neural network-based methods is a divergence with the ground truth in the high-wavenumber region of the spectrum. This suggests that the models face challenges in accurately capturing the smallest scale features.

\subsubsection{Wall pressure fluctuation of turbulent boundary layers}
Instantaneous wall-pressure fluctuations $p'(x,z,t)$ in zero-pressure-gradient turbulent boundary layers are dominated by fine, intermittent structures generated by near-wall vortical events and their footprints; the field is broadband in both space and time and more challenging than the inlet-velocity slices considered earlier. This makes the problem a stringent test of whether the models can capture locally varying dynamics.

\begin{figure}[t!]
    \centering
    \includegraphics[width=0.85\linewidth]{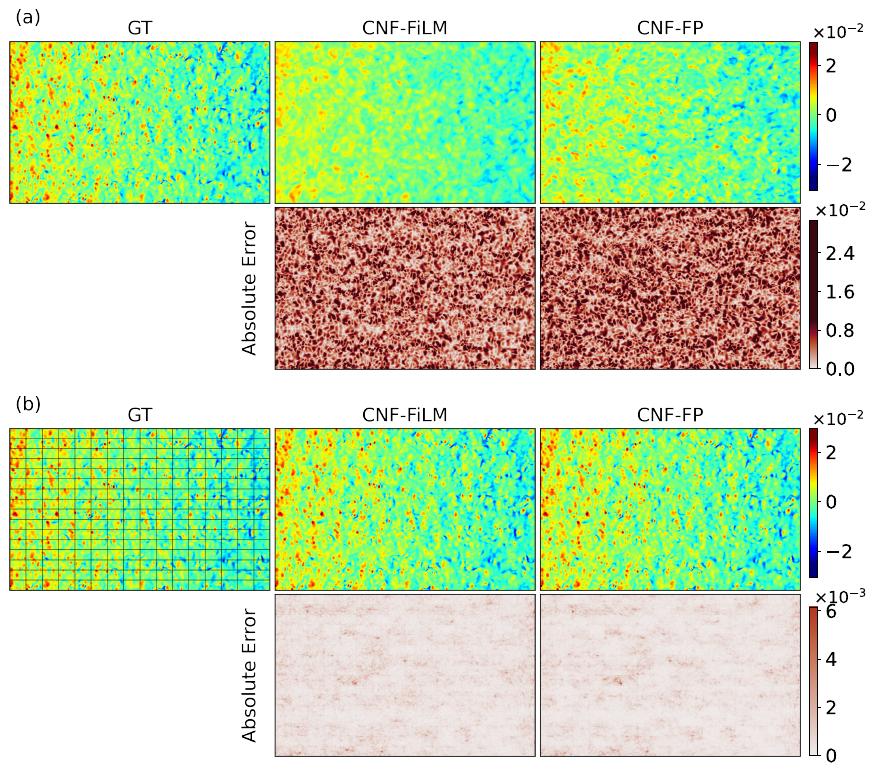}
    \caption{Wall-pressure fluctuations for \emph{out-of-range testing} at $r=128$. Top: reconstructions; bottom: absolute error. (a) Global CNFs (w/o DD) blur intermittent high–amplitude regions; errors are large and structured (scale $\times10^{-2}$). (b) Domain-decomposed CNFs recover the spatial distribution and peaks; errors are diffuse and about an order of magnitude smaller (scale $\times10^{-3}$).}
    \label{fig:snapshot_extrap_pressure}
\end{figure}

Figure~\ref{fig:snapshot_extrap_pressure} compares out-of-range testing results at $r{=}128$ with and without domain decomposition (DD). Global CNFs (w/o DD) reproduce the broad streamwise pattern but smear and misplace intermittent high–amplitude regions; the accompanying absolute–error maps exhibit large, structured residuals that persist across the field, indicating both amplitude and phase errors. With DD, both CNF--FiLM and CNF--FP recover the spatial distribution and amplitudes of $p'$ much more faithfully, and the error maps become diffuse and an order of magnitude smaller, indicating localization suppresses the over-smoothing and misalignment inherent to a single global latent on this intermittent signal.

Table~\ref{tab:pressure_error} summarizes relative $L^2$ errors across latent sizes. DD yields large improvements at every $r$: at $r{=}128$, CNF--FP improves from $24.45\% \to 8.92\%$ (training) and from $109.44\% \to 8.72\%$ (out-of-range testing); CNF--FiLM improves from $78.89\% \to 10.19\%$ (training) and from $99.81\% \to 9.49\%$ (out-of-range testing). Similar trends hold at $r{=}64$ and $r{=}32$. Without DD, both conditionings fail dramatically out of range (errors $> 100\%$), confirming that a single global latent is inadequate for these complex pressure fluctuation fields. With DD, CNF--FP typically fits training best, whereas FiLM and FP are comparable on extrapolation at large $r$.
\begin{table}[t!]\centering
\scriptsize
\begin{tabular}{l l c c c c}
\toprule
& & \multicolumn{2}{c}{\textbf{With Domain Decomposition}} & \multicolumn{2}{c}{\textbf{Without Domain Decomposition}} \\
\cmidrule(lr){3-4}\cmidrule(lr){5-6}
\textbf{Latent size} & \textbf{Dataset} & \textbf{CNF-FP} & \textbf{CNF-FiLM} & \textbf{CNF-FP} & \textbf{CNF-FiLM} \\
\midrule
\multirow{2}{*}{32}
& Training & \textbf{32.08\%} & 40.94\% & 46.43\% & 85.53\% \\
& Testing extrap & 41.53\% & \textbf{39.61\%} & 128.43\% & 104.01\% \\
\midrule
\multirow{2}{*}{64}
    & Training &  \textbf{14.67\%} & 23.50\% & 34.03\% & 82.70\% \\
& Testing extrap & \textbf{17.83\%} & 22.82\% & 122.78\% & 101.71\% \\
\midrule
\multirow{2}{*}{128}
& Training & \textbf{8.92\%} & 10.19\% & 24.45\% & 78.89\% \\
& Testing extrap &\textbf{ 8.72\%} & 9.49\% & 109.44\% & 99.81\% \\
\bottomrule
\end{tabular}
\caption{Relative $L^2$ errors (\%) on the wall pressure fluctuation dataset, comparing CNFs \emph{with} and \emph{without} domain decompositions.}
\label{tab:pressure_error}
\end{table}

\begin{figure}[t!]
    \centering
    \includegraphics[width=1\linewidth]{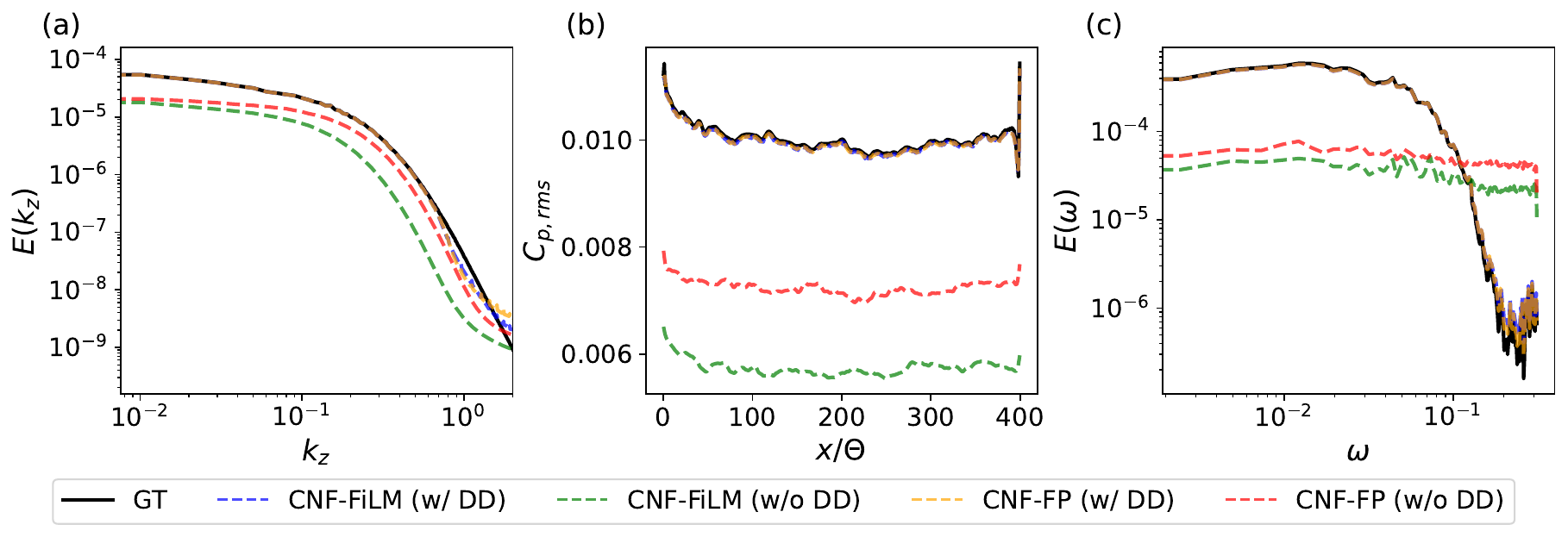}
    \caption{Statistics of wall-bounded pressure fluctuations, out-of-range testing at latent size $r = 128$. (a) Spanwise spectrum $E(k_z)$; (b) RMS of $p'$ versus normalized streamwise location $x/\Theta$; (c) Frequency spectrum $E(\omega)$ at a centerline probe.}
    \label{fig:pressure_stats_extrap}
\end{figure}

Figure~\ref{fig:pressure_stats_extrap} (out-of-range testing, $r{=}128$) evaluates spatial and temporal statistics of the reconstructed wall-pressure field.  Panel~(a) shows the spanwise spectrum $E(k_z)$: models with DD track the DNS closely across the inertial and dissipative ranges, delaying spectral roll-off to higher $k_z$. In contrast, global CNFs (w/o DD) exhibit a broadband energy deficit, most severe for CNF--FiLM (green), with premature decay at moderate $k_z$; CNF--FP (red) is less extreme but still biased low. Panel~(b) reports the RMS of $p'$ versus the normalized streamwise coordinate $x/\Theta$: DD preserves both the magnitude and the weak streamwise variation of the intensity, whereas global CNFs remain uniformly underpowered along $x$. Panel~(c) presents the frequency spectrum $E(\omega)$ at a centerline probe: DD reproduces the broadband shape and cutoff frequency, while global CNFs show an elevated low-frequency plateau and insufficient decay at high $\omega$, which is completely off from the reference. 
Taken together, these diagnostics show that domain decomposition is necessary to recover the spatial and temporal spectral content of intermittent wall-pressure fluctuations under out-of-range testing.

\section{Discussion}
\label{sec:Discussion}

\subsection{Interpreting learned CNF modes and latent geometry}
A CNF decodes a low–dimensional latent $\mathbf{z}\in\mathbb{R}^r$ into a continuous field by modulating a shared base decoder. Intuitively, a CNF mode is the spatial pattern that appears in the output field when one moves in a particular direction of the latent space. From this idea, we define a finite–amplitude mode field, which shows the nonlinear change in the reconstructed field produced by exciting one latent direction to a standardized amplitude. Let the trained decoder be
\begin{equation}
\bm{\phi}(\mathbf{x};\mathbf{z}) =
f\big(\mathbf{x}; \boldsymbol{\theta}^* + \mathbf{M}^* \mathbf{z}\big),
\qquad
\mathbf{z}\in\mathbb{R}^{r},
\label{eq:cnf_decode_disc}
\end{equation}
where $\boldsymbol{\theta}^*$ are trained base parameters and $M^* \in \mathbb{R}^{h\times r}$ maps the latent to a parameter update. The zero CNF mode $\bm{\phi}^{(0)}$ is defined as the unconditioned output, which is obtained by setting the latent vector to zero:
\begin{equation}
    \bm{\phi}^{(0)} := f(\mathbf{x}; \boldsymbol{\theta}^*)
\end{equation}
For the $i^{th}$ latent axis, we define the mode–excited field at standardized amplitude $\alpha > 0$ as
\begin{equation}
\bm{\phi}^{(i)}(\mathbf{x}; \alpha) := f \big(\mathbf{x}; \bm{\theta}^* + \mathbf{M}^*\alpha\mathbf{e}_i\big),
\qquad i=1,\ldots,r,
\end{equation}
where $\bm{e}_i$ is the unit basis vector and the corresponding mode increment as
\begin{equation}
\Delta\bm{\phi}^{(i)}(\mathbf{x};\alpha):= \bm{\phi}^{(i)}(\mathbf{x};\alpha) - \bm{\phi}^{(0)}(\mathbf{x}).
\label{eq:finite_mode_increment}
\end{equation}
In practice we choose $\alpha$ to make different directions comparable, e.g. $\alpha=1$ or 
$\alpha=\sigma_i$ (the empirical standard deviation of the $i^{th}$ latent over the training set.

\begin{figure}[!thb]
    \centering
    \includegraphics[width=1\linewidth]{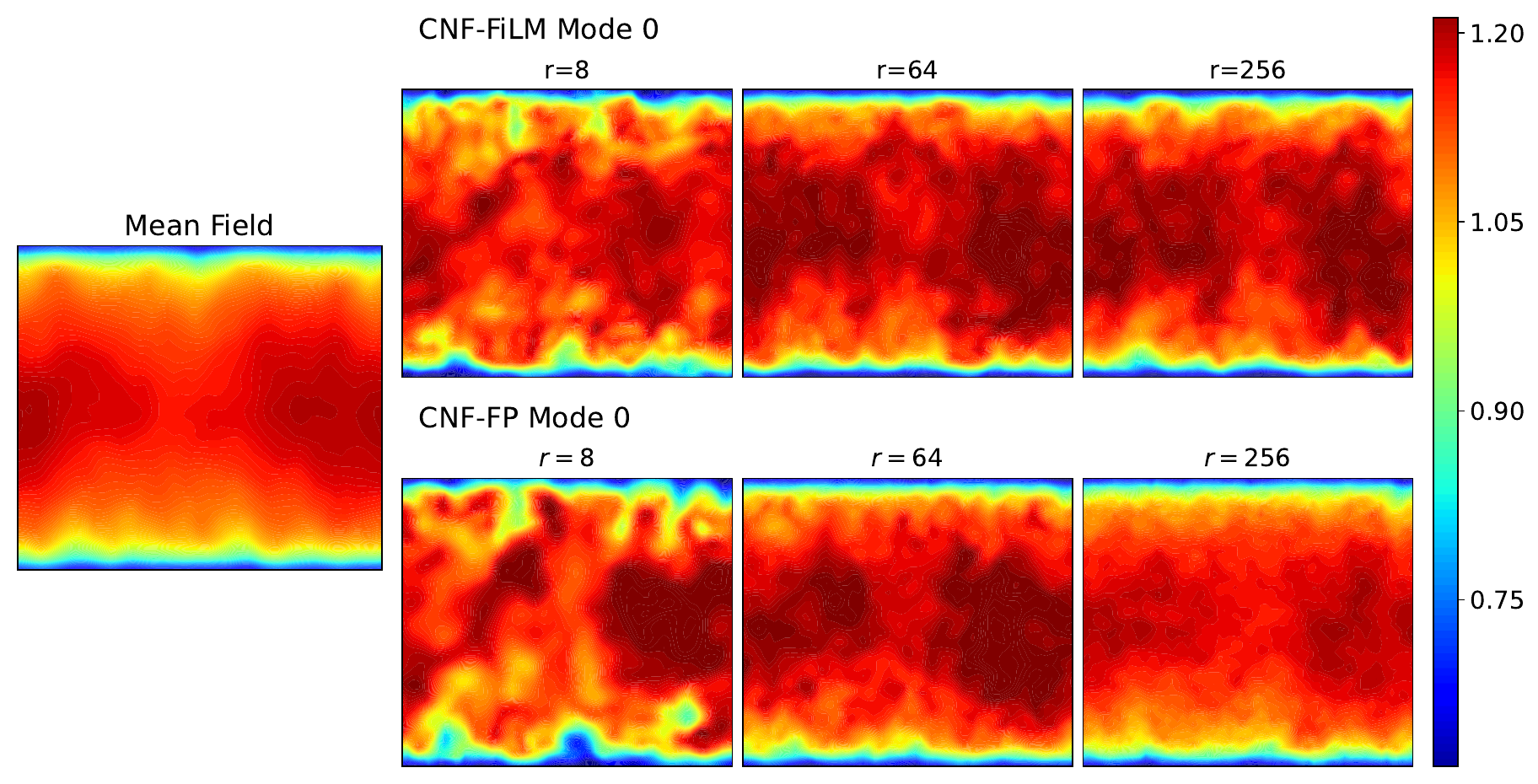}
    \caption{Mean field and CNF mode visualization at different latent size configuration (8, 64 256) for WMLES-Inlet dataset }
    \label{fig:mode_0_compare}
\end{figure}

Figure~\ref{fig:mode_0_compare} compares the empirical mean field with the CNF base field (``Mode~0'') for FiLM and FP conditioning across $r\in\{8,64,256\}$ on the WMLES–Inlet dataset. 
In all cases, Mode~0 reproduces the large-scale organization of the mean (centerlilne high, near-wall low) for all settings.  
For FiLM, Mode~0 is already close to the mean at $r=8$ and changes mildly as $r$ increases; differences are localized near the walls and along gentle large-scale undulations, consistent with a base decoder that carries the low-frequency ``average'' structure while the latent biases modulate departures. For FP, discrepancies are more pronounced at low latent size ($r = 8$); as $r$ grows to $64$ and $256$, Mode~0 progressively approaches the mean and the large–scale bias diminishes. These trends hold true regardless of hyperparameters and training recipes.

\begin{figure}[t!]
    \centering
    \includegraphics[width=0.8\linewidth]{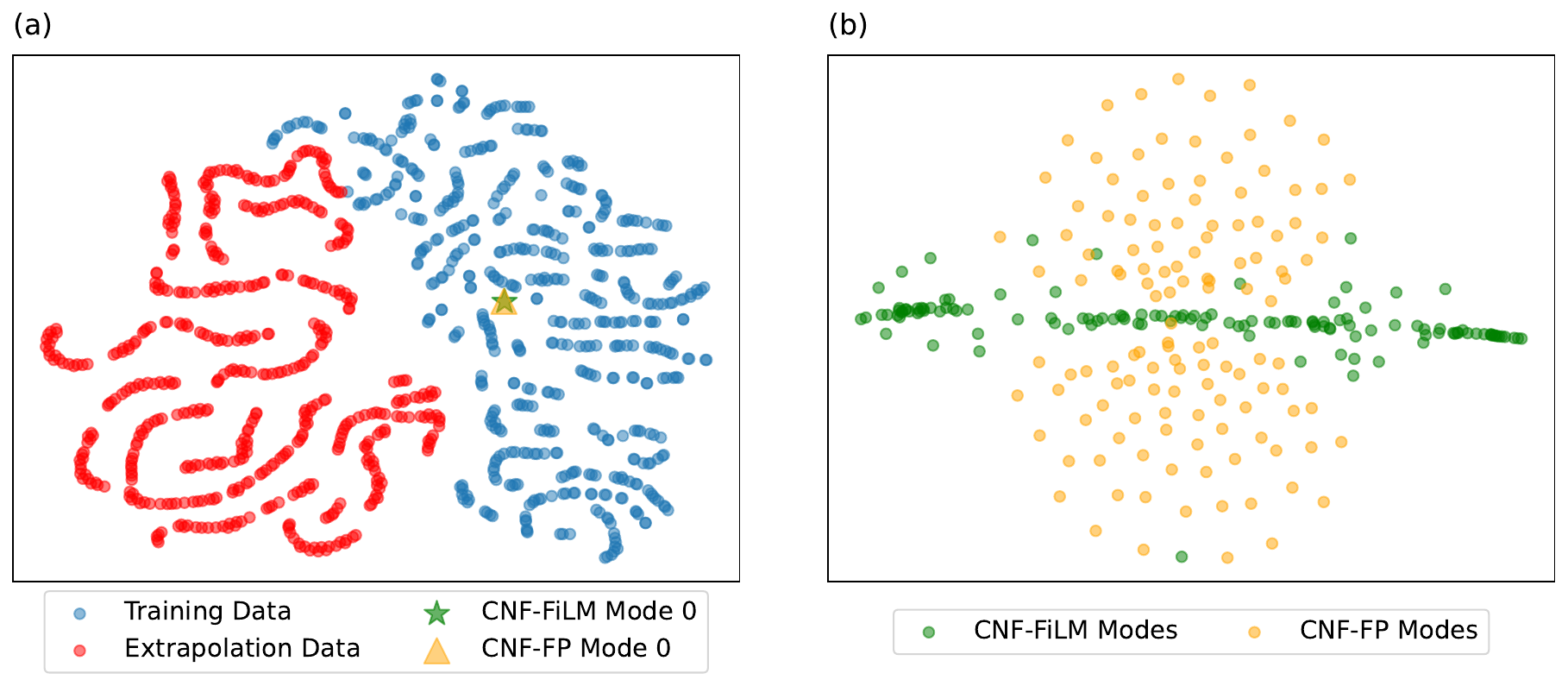}
    \caption{(a) T-SNE analysis on training WMLES-Inlet data. (b) T-SNE analysis of CNF modes}
    \label{fig:mode_train_extrap_tsne}
\end{figure}

Figure~\ref{fig:mode_train_extrap_tsne} provides a qualitative view of the learned latent geometry using t-SNE (a non-linear embedding that preserves local neighborhoods but distorts global distances; we therefore use it only for visualization). 
In panel~(a), the training snapshots (blue) form a coherent manifold that is clearly separated from the out-of-range testing snapshots (red), consistent with our split design in Sec.~\ref{subsec:Experiment_setting}. The Mode~0 markers for FiLM (green star) and FP (orange triangle) lie near the centroid of the training cloud, in line with Fig.~\ref{fig:mode_0_compare} and the interpretation of the unconditioned decoder as a mean–like representative of the training distribution. Panel~(b) embeds \emph{finite–amplitude mode} fields generated by unit excitations along single latent axes. The FiLM modes occupy a narrow, nearly one–dimensional band, whereas the FP modes spread over a substantially larger area of the embedding. This broader dispersion indicates that individual FP coordinates induce a wider variety of field perturbations than FiLM coordinates, which is consistent with FP's stronger fitting ability and larger out-of-range generalization gap reported in Sec.~\ref{sec:Result}. We emphasize that t-SNE does not support metric claims; a principled, basis-aware quantification can be obtained by analyzing the energy and rank of the standardized mode increments $\{\Delta\bm{\phi}^{(i)}(\cdot;\alpha)\}$, which we view as complementary future diagnostics.

%The t-SNE plot in Figure~\ref{fig:mode_train_extrap_tsne} further supports these observations and provides additional insights into the models' representational capacities. In Figure~\ref{fig:mode_train_extrap_tsne}a, Mode 0 is clearly positioned at the center of the training data cluster, confirming its role as a central representation of the learned distribution. Concurrently, the extrapolation data points are distinctly located outside the main cluster of training data, highlighting their out-of-distribution nature. Figure~\ref{fig:mode_train_extrap_tsne}b offers an explanation for the superior fitting capabilities of the CNF-FP model compared to CNF-FiLM. The CNF-FP modes span a noticeably larger space on the t-SNE manifold than those of CNF-FiLM models. This expanded span indicates CNF-FP's capacity to adapt to a wider range of features and capture more intricate details within the training data. Although this augmented representational power benefits fitting, it appears to hinder generalization to unseen data, resulting in higher extrapolation errors for CNF-FP compared to CNF-FiLM (as previously observed). This outcome underscores a trade-off between model expressiveness and generalization performance in these architectures. 

\subsection{Conditioning strategy, capacity, and generalization}
\label{sec:discussion_on_generalization}
We probed whether the generalization gap stems from how expressive power is placed in the architecture rather than from raw trainable parameter count. First, we fixed an identical base decoder $f(\cdot;\theta)$ (depth, widths, activations) for both mechanisms and trained across latent sizes $r$. Under this control, FP (which applies latent–driven {weight and bias} updates $\Delta W^{(\ell)}(\mathbf{z}),\Delta b^{(\ell)}(\mathbf{z})$) consistently achieved lower training and in–range errors but exhibited a larger extrapolation error $\Delta_{\text{gen}} := \mathcal{E}_{\text{extrap}}-\mathcal{E}_{\text{train}}$ than FiLM (which modulates {biases only}). Second, to rule out parameter-count effects, we matched the \emph{total} number of trainable parameters by increasing FiLM's base widths. The pattern persisted across datasets and $r$: FP remained the best fitter yet generalized worse under strict out-of-range testing, whereas FiLM maintained smaller $\Delta_{\text{gen}}$. The details of the result are presented in~\ref{sec:sup_gen_discussion}.

A mechanistic explanation follows from the latent-to-output sensitivity. 
Let $J_z(\mathbf{x};\mathbf{z})=\partial \bm{\phi}(\mathbf{x};\mathbf{z})/\partial\mathbf{z}$ denote the Jacobian of the decoder with respect to the latent, its size $\|J_z\|$ measures the gain from latent perturbations to field changes. Under matched parameter budgets, FP’s multiplicative weight modulation $\Delta W^{(\ell)}(\mathbf{z})$ yields a larger gain than FiLM’s additive bias modulation $\Delta b^{(\ell)}(\mathbf{z})$, enabling sharper snapshot-specific adaptations (lower training/in-range error) but amplifying errors in out-of-range tests. Practically, capacity should therefore be allocated rather than merely increased: prefer FiLM when extrapolation is critical or combine either mechanism with domain decomposition to localize complexity; use FP when the priority is best possible in-distribution accuracy. 
%In our main experiments, we first ensured that the base networks used for FiLM and FP modulation contained the same number of parameters. Under this setting, we observed that FP, despite achieving higher fitting accuracy, was more prone to overfitting and exhibited weaker generalization than FiLM. One might argue that this outcome is driven by a discrepancy in the total number of parameters, since FiLM typically introduces fewer additional parameters than FP. To address this, we conducted experiments where the total trainable parameter counts of the two strategies were matched. The results remained consistent: FP continued to suffer from overfitting, whereas FiLM maintained stronger generalization performance.

\section{Conclusion}
We presented a unified encoding–decoding framework to benchmark spatial dimensionality reduction methods for turbulent flows, placing CNFs alongside POD and CNN-AEs under identical preprocessing, metrics, and fixed train/interpolation/extrapolation splits. In contrast to most prior studies that evaluate only interpolative testing accuracy, our protocol explicitly separates \emph{in-range} from \emph{strict out-of-range} testing and augments pointwise errors with physics-grounded diagnostics (turbulence statistics measures).

First, among learnable decoders, CNF with full projected weight and bias modulation (CNF–FP) delivers the strongest data fitting and in-range reconstruction across latent sizes, whereas activation-only modulation (CNF–FiLM) generalizes more reliably under strict extrapolation once moderate capacity is available; linear POD is competitive in $L^2$ only at very small latent dimension and underrecovers fluctuation statistics. Second, when flows become more demanding, a single global latent is insufficient; a domain-decomposed CNF that localizes the mapping markedly improves extrapolation accuracy and better preserves near-wall peaks and high-wavenumber content. Third, analysis of CNF ``modes'' and latent–to–output sensitivity provides a mechanism for these trends: weight modulation increases latent gain, aiding fit but amplifying errors under distribution shift, whereas bias-only modulation yields a lower, more uniform gain and thus smaller extrapolation gaps. These results lead to practical guidance. For applications that prioritize robustness beyond the training horizon, prefer CNF–FiLM and allocate capacity spatially via domain decomposition; for best in-range accuracy, CNF–FP is effective, provided latent sensitivity is controlled. 

Limitations and opportunities remain. Our study focuses on spatial reduction with auto-decoding; future work should assess amortized encoders for fast inference under partial observations, extend domain decomposition with overlap and adaptive tiling, and couple spatial CNFs with temporal models for fully spatiotemporal reduction. Incorporating uncertainty quantification for latents and sensitivity-aware training objectives may further stabilize extrapolation. We expect the evaluation protocol and analyses here to serve as a physics-aware basis for choosing conditioning, capacity, and localization when deploying CNFs for turbulence compression, reconstruction, and as building blocks for operator learning and generative flow models.

\section*{Acknowledgements}
 The authors would like to acknowledge the funds from Office of Naval Research under award numbers N00014-23-1-2071 and National Science Foundation under award numbers OAC-2047127.

\clearpage
\appendix

\section{Dataset generation}
\label{app:data}
We construct training and testing datasets from two benchmark flow configurations: a 3D turbulent channel flow and a 3D turbulent flat boundary layer. Both cases follow the unsteady incompressible Navier--Stokes equations:

\begin{equation}
    \begin{aligned}
    \frac{\partial \mathbf{u}}{\partial t}+(\mathbf{u} \cdot \nabla) \mathbf{u} & =-\nabla p+\nu \nabla^2 \mathbf{u}+\mathbf{f}, \\
    \nabla \cdot \mathbf{u} & =0,
    \end{aligned}
\end{equation}
where $\mathbf{u}(\mathbf{x}, t)$ denotes the velocity vector, $p(\mathbf{x}, t)$ the pressure, $\nu$ the viscosity, and $\mathbf{f}(\mathbf{x}, t)$ the forcing term. We employ a wall-modeled large-eddy simulation (WMLES) framework and direct numerical simulation (DNS) within our in-house Navier-Stokes solver to generate datasets~\cite{fan2025diff}. 
% The schematic of dataset extraction can be seen from Figure~\ref{fig:3Dschemetic}.
\begin{figure}[!h]
    \centering
    \includegraphics[width=0.95\linewidth]{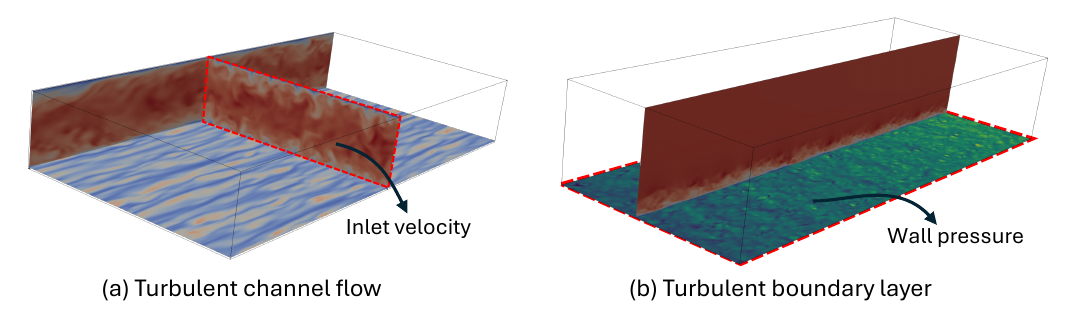}
    \caption{The schematic of dataset extraction from 3D simulations. We crop  the data from the 2D plane marked in red dashed line.}
    \label{fig:3Dschemetic}
\end{figure}
\subsection{Turbulent channel flow (WMLES)} 
The friction Reynolds number is $Re_\tau = 6000$. The computational domain measures $2 \pi \times 2 \times \pi$ and is discretized using a structured grid of $64 \times 64 \times 64$. We advance the solution with a uniform time step $\Delta t=0.05 \mathrm{~s}$. For training and evaluation, we consider only the velocity field in the streamwise direction ($x$) and extract two-dimensional inlet plane snapshots with a resolution of $64 \times 64$. We collect 3000 total snapshots from $t=0 \mathrm{~s}$ to $t=150 \mathrm{~s}$. From these, 1000 randomly chosen snapshots in $t \in [0,100)$ serve as training data, with the remaining snapshots in this interval used for interpolation testing. All snapshots in $t \in [100,150)$ are reserved for extrapolation testing.

\subsection{Turbulent channel flow (DNS)}
The simulations are performed at a friction Reynolds number of $Re_\tau = 180$. The computational domain has dimensions $4\pi \times 2 \times 2\pi$ and is discretized on a structured grid of $320 \times 400 \times 200$ points. Time integration is carried out with a uniform step size of $\Delta t = 5 \times 10^{-4}\,\mathrm{s}$. For training and evaluation, we focus on the streamwise velocity field and extract two-dimensional inlet-plane snapshots with a resolution of $400 \times 200$. A total of 21,900 snapshots are collected over the interval $t \in [0,1095]$, sampled every 100 numerical steps. From these, 1,000 randomly selected snapshots in $t \in [0,100)$ are used for training, while the remaining snapshots in this interval are employed for interpolation testing. All snapshots in $[100,150)$ are reserved for extrapolation testing.  

\subsection{Turbulent flat boundary layer (DNS)}
 The Reynolds number based on the free-stream velocity and inlet momentum thickness is $Re_{\boldsymbol{\theta}} = 300$. The computational domain, normalized by the inlet momentum thickness, is $L_x \times L_y \times L_z = 400 \times 80 \times 160$. The domain is uniformly discretized in the streamwise ($x$) and spanwise ($z$) directions, while a stretched grid is employed in the wall-normal ($y$) direction, with a total resolution of $N_x \times N_y \times N_z = 512 \times 320 \times 512$. Wall pressure fluctuation data are sampled at intervals of $\Delta T =1.6$ from $x-z$ plane, which corresponds to ten times the numerical time step. The total duration of the collected wall pressure data spans 12 flow-through times, yielding 3000 snapshots. Of these, the first 2000 snapshots are used as the training and interpolation dataset, while the last 1000 snapshots are reserved for extrapolation testing.

% \jg{@xiantao please help me write here,data shape is (21900, 1, 400, 200),data path you give me is 
% \path{/home/xiantao/case/wall_pressure/inlet/inletdata/channel_180_u_x200_41-259.npy}
% }

\section{Model implementation details}

\subsection{CNF \& decomposed CNF architecture and hyper-parameters}

As outlined in the methodology, our CNF model's base network maps input coordinates to field values. We implement this network using a Multilayer Perceptron (MLP) based on the SIREN architecture \cite{sitzmann2020implicit}, which employs $\omega_0$-scaled sine activation functions, $\sigma(x)=\sin \left(\omega_0 x\right)$, for all hidden layers. This choice is motivated by SIREN's effectiveness in representing continuous signals and their derivatives, crucial for accurately capturing field details and mitigating the spectral bias towards low frequencies often seen in standard MLPs \cite{rahaman2019spectral}. Our specific architecture uses 8 hidden layers, each with 64 neurons ($n_h=64$) and  frequency parameter $\omega_0=30$ for all testing cases. Furthermore, SIREN requires a specific weight initialization scheme. For the first layer, weights $W^{(0)}$ are initialized element-wise from a uniform distribution $\mathcal{U}\left(-1 / n_i, 1 / n_i\right)$. For all subsequent hidden layers, weights $\{W^{(l)}\}_{l=1}^{8}$ are initialized element-wise from $\mathcal{U}\left(-\frac{\sqrt{6 / n_h}}{\omega_0}, \frac{\sqrt{6 / n_h}}{\omega_0}\right)$. 

For the decomposed framework, it was observed that optimal performance necessitates the use of distinct patch sizes for different datasets. Specifically, the $512 \times 512$ full domain of the Wall-Pressure fluctuation dataset is decomposed into $32 \times 32$ non-overlapping patches, while the $400 \times 200$ full domain of the Inlet-DNS dataset is decomposed into $20 \times 20$ non-overlapping patches.

\subsection{Baseline CNN-AE architecture and hyper-parameter selection}

\begin{table}[h]
  \centering
    \resizebox{\textwidth}{!}{ 
  \begin{tabular}{@{}lcccccc@{}}
    \toprule
    Layer Type                          & \multicolumn{6}{c}{Latent size} \\
    \cmidrule(lr){2-7}
                                    & 8                   & 16                  & 32                  & 64                  & 128                 & 256                 \\
    \midrule
    % Encoder
    Conv2D             & (B, 31, 32, 32)    & (B, 37, 32, 32)    & (B, 42, 32, 32)    & (B, 52, 32, 32)    & (B, 54, 32, 32)    & (B, 60, 32, 32)    \\
    BatchNorm2d                    & (B, 31, 32, 32)    & (B, 37, 32, 32)    & (B, 42, 32, 32)    & (B, 52, 32, 32)    & (B, 54, 32, 32)    & (B, 60, 32, 32)    \\
    ReLU                           & (B, 31, 32, 32)    & (B, 37, 32, 32)    & (B, 42, 32, 32)    & (B, 52, 32, 32)    & (B, 54, 32, 32)    & (B, 60, 32, 32)    \\
    Conv2D              & (B, 62, 16, 16)    & (B, 74, 16, 16)    & (B, 84, 16, 16)    & (B,104, 16, 16)    & (B,108, 16, 16)    & (B,120, 16, 16)    \\
    BatchNorm2d                   & (B, 62, 16, 16)    & (B, 74, 16, 16)    & (B, 84, 16, 16)    & (B,104, 16, 16)    & (B,108, 16, 16)    & (B,120, 16, 16)    \\
    ReLU                           & (B, 62, 16, 16)    & (B, 74, 16, 16)    & (B, 84, 16, 16)    & (B,104, 16, 16)    & (B,108, 16, 16)    & (B,120, 16, 16)    \\
    Conv2D                           & (B,124,  8,  8)    & (B,148,  8,  8)    & (B,168,  8,  8)    & (B,208,  8,  8)    & (B,216,  8,  8)    & (B,240,  8,  8)    \\
    BatchNorm2d                    & (B,124,  8,  8)    & (B,148,  8,  8)    & (B,168,  8,  8)    & (B,208,  8,  8)    & (B,216,  8,  8)    & (B,240,  8,  8)    \\
    ReLU                           & (B,124,  8,  8)    & (B,148,  8,  8)    & (B,168,  8,  8)    & (B,208,  8,  8)    & (B,216,  8,  8)    & (B,240,  8,  8)    \\
    \midrule
    Flatten                         & (B, 7936)          & (B, 9472)          & (B,10752)          & (B,13312)          & (B,13824)          & (B,15360)          \\
    Linear                  & (B,    8)          & (B,   16)          & (B,   32)          & (B,   64)          & (B,  128)          & (B,  256)          \\
    Linear                  & (B, 7936)          & (B, 9472)          & (B,10752)          & (B,13312)          & (B,13824)          & (B,15360)          \\
    Unflatten                       & (B,124,  8,  8)    & (B,148,  8,  8)    & (B,168,  8,  8)    & (B,208,  8,  8)    & (B,216,  8,  8)    & (B,240,  8,  8)    \\
    \midrule
    ConvTranspose2d           & (B, 62, 16, 16)    & (B, 74, 16, 16)    & (B, 84, 16, 16)    & (B,104, 16, 16)    & (B,108, 16, 16)    & (B,120, 16, 16)    \\
    BatchNorm2d                   & (B, 62, 16, 16)    & (B, 74, 16, 16)    & (B, 84, 16, 16)    & (B,104, 16, 16)    & (B,108, 16, 16)    & (B,120, 16, 16)    \\
    ReLU                          & (B, 62, 16, 16)    & (B, 74, 16, 16)    & (B, 84, 16, 16)    & (B,104, 16, 16)    & (B,108, 16, 16)    & (B,120, 16, 16)    \\
    ConvTranspose2d           & (B, 31, 32, 32)    & (B, 37, 32, 32)    & (B, 42, 32, 32)    & (B, 52, 32, 32)    & (B, 54, 32, 32)    & (B, 60, 32, 32)    \\
    BatchNorm2d                   & (B, 31, 32, 32)    & (B, 37, 32, 32)    & (B, 42, 32, 32)    & (B, 52, 32, 32)    & (B, 54, 32, 32)    & (B, 60, 32, 32)    \\
    ReLU                         & (B, 31, 32, 32)    & (B, 37, 32, 32)    & (B, 42, 32, 32)    & (B, 52, 32, 32)    & (B, 54, 32, 32)    & (B, 60, 32, 32)    \\
    ConvTranspose2d         & (B,1,64,64) & (B,1,64,64) & (B,1,64,64) & (B,1,64,64) & (B,1,64,64) & (B,1,64,64) \\
    \bottomrule
    total params & 309389 & 560789 & 1018449 & 2206529 & 4080369 & 8530817 \\
    \bottomrule
  \end{tabular}
  }
    \caption{CNN-AE output shapes by layer for different latent size configuration}
    \label{tab:CNN_AE_config}
\end{table}

For baseline comparison, we implemented a standard convolutional autoencoder (CNN‑AE) closely following Pan et al. \cite{pan2023neural}. Its encoder consists of three convolutional stages—each a $3\times3$ kernel, stride 2, padding 1 convolution followed by batch normalization and ReLU—and a fully connected layer projecting to the latent space. The decoder mirrors this design with three $3\times3$ transposed convolutions (stride 2, padding 1, padding 1), each again paired with batch normalization and ReLU, to recover the original spatial resolution. To ensure fair comparison in representational capacity, we tuned the number of channels at every stage so that, for each latent size configuration, the total parameter count of the CNN‑AE matches that of our CNF model. Detailed per‑layer output shapes and trainable parameter counts for all latent‑size configurations appear in Table~\ref{tab:CNN_AE_config}.

\section{Evaluation Metrics}
\paragraph{Relative $L^2$ error} To quantify the model's performance, we adopt the mean relative $L^2$ error. Let $\{\boldsymbol{\phi}^i\}_{i=1}^{N}$ denote the ensemble of the ground truth turbulence data used in this work, and $\{\boldsymbol{\hat{\phi}}^i\}_{i=1}^{N}$ denote the corresponding model predictions. Relative $L^2$ error $\epsilon$ is defined as:
    \begin{equation}
        \epsilon = \frac{1}{N}\sum_{i=1}^N\frac{\|\boldsymbol{\phi}^i-\boldsymbol{\hat{\phi}}^i\|_2}{\|\boldsymbol{\phi}^i\|_2},
    \end{equation}
    where $\|\cdot\|_2$ denotes the standard Euclidean norm. 
\paragraph{Turbulence statistics} 
To compute the turbulence statistics, we first decompose flow quantities $q(x, y, z, t)$  into a mean $\langle q \rangle(y)$ and fluctuating $q^\prime$ components, i.e. $q^\prime = q -\langle q \rangle$. Here, $\langle \cdot \rangle$ denotes the the operator that averages over time and the homogeneous directions. 
All the turbulence statistics are normalized by the inner scale, taking the friction velocity $u_\tau$ as the reference. Spatial dimensions are normalized by viscous length, $\delta_\nu=\nu / u_\tau$. For example, the dimensionless wall-normal coordinate is defined as $y^{+}=y / \delta_\nu$.

The normalized root-mean-square of velocity fluctuation $C_{u,rms}$ is computed as: 
\begin{equation}
    C_{u,rms} = \frac{\sqrt{\langle u^{\prime2}\rangle}}{u_{\tau}}.
\end{equation}
For wall pressure measured at $y=0$, with fluctuation $p^\prime$, the normalized root-mean-square is:
\begin{equation}
    C_{p,rms} = \frac{\sqrt{\langle p^{\prime2}\rangle}}{\rho u_{\tau}^2}.
\end{equation}
We report 1D spectrum consistent with an energy-conserving FFT implementation. Define two-sided power spectral density (PSD) $\Psi(\omega,k_x,k_z)$ of $q^\prime$. The 1D spanwise wavenumber spectrum is obtained by integrating out frequency and streamwise wavenumber and then adopting a one-sided convention in $k_z >0$:
\begin{equation}
E\left(k_z\right)=\left\langle\int_0^{\infty} \int_0^{\infty} \Psi\left(\omega, k_x, k_z\right) d k_x d \omega\right\rangle.
\end{equation} 
The 1D temporal spectrum at a fixed streamwise location $x_{loc}$ is the temporal PSD of $q^{\prime}(t, x_{loc}, z)$ averaged over $z$, denoted $S(\omega ; x_{loc})$, with
\begin{equation}
        E(\omega) = \left\langle\int_0^{\infty} S\left(\omega ; x_{loc}\right) d \omega \right\rangle.
\end{equation}

\section{Additional results}
\label{app:equal_param_results}
This section presents supplementary results supporting the main paper's findings. We provide a visualization of the flow field reconstructions referenced in Sec.~\ref{sec:benchmark_study} and quantitative results of the study discussed in Sec.~\ref{sec:discussion_on_generalization}. 
\subsection{Supplementary visualization for benchmark study}
\begin{figure}[!thb]
    \centering
    \includegraphics[width=1\linewidth]{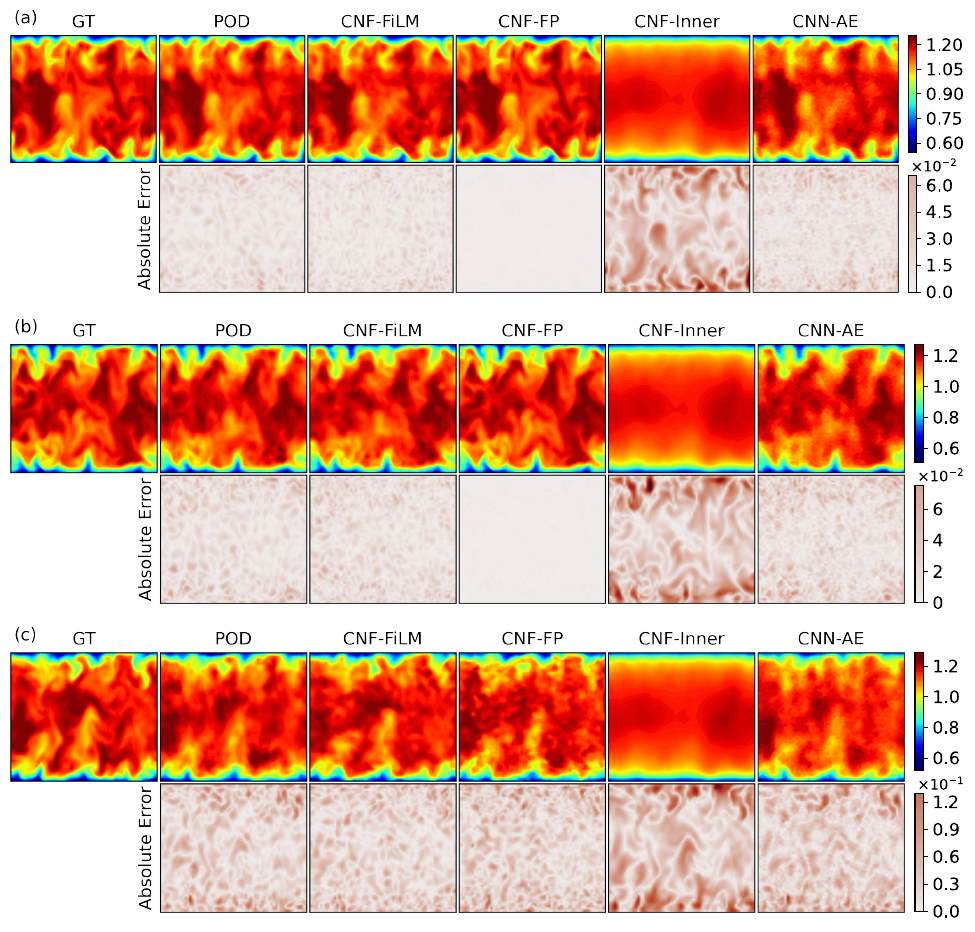}
    \caption{WMLES--Inlet snapshots at latent size $r=128$. Each panel shows the reconstructed streamwise–velocity field (top) and the corresponding absolute error (bottom). Rows follow our evaluation splits: (a) training, (b) in-range testing, (c) out-of-range testing.}
    \label{fig:snapshot_LES_Inlet_combined}
\end{figure}
Figure~\ref{fig:snapshot_LES_Inlet_combined} offers a qualitative assessment of the model's reconstruction fidelity for a latent size of $r=128$. The reconstructions for CNF-FP for the training and in-range test cases are visually indistinguishable from the ground truth, with absolute errors that are small in magnitude. This indicates the CNF-FP's strong capacity for high-fidelity representation of data within the training distribution. For the out-of-range test case, while the primary flow structures are accurately captured, a noticeable increase in reconstruction error is observed. These errors appear to be concentrated in regions characterized by high spatial frequency content and strong velocity gradients, which are expected challenges when generalizing to unseen flow conditions. 

\subsection{Supplementary result for generalization discussion}
\label{sec:sup_gen_discussion}
We construct a new baseline model, denoted CNF-FiLM*, specifically designed to have the same total number of trainable parameters as the FP model. To achieve this parameter equivalence, we increase its hidden layer width while keeping its depth fixed. Table~\ref{tab:cnf_params} lists the architectural choices and resulting parameter counts; reconstruction errors for training, interpolation, and extrapolation splits are given in Table~\ref{tab:cnf_errors_equal_param}.

The CNF-FP model demonstrates superior performance on both training and interpolation splits, achieving the lowest error. This suggests its architectural design is more efficient for fitting in-distribution data. For out-of-distribution data, the most compact model CNF-FiLM achieves the lowest error. The proposed CNF-FP is the second-best performer, whereas the high-capacity CNF-FiLM* exhibits the worst generalization. This outcome indicates that over-parameterizing the FiLM architecture is detrimental to its extrapolation capabilities. 

\begin{table}[!h]
\centering
\begin{tabular}{lcccc}
\toprule
Latent size & Configuration & CNF-FiLM* & CNF-FP & CNF-FiLM \\
\midrule
\multirow{3}{*}{32} & Network width & 350 & 64 & 64 \\
                    & Network depth & 8 & 8 & 8 \\
                    & Total trainable parameters & $1.09\times10^{6}$ & $1.10\times10^{6}$ & $5.20\times10^{4}$ \\
\midrule
\multirow{3}{*}{64} & Network width & 480 & 64 & 64 \\
                    & Network depth & 8 & 8 & 8 \\
                    & Total trainable parameters & $2.13\times10^{6}$ & $2.18\times10^{6}$ & $7.04\times10^{4}$ \\
\midrule
\multirow{3}{*}{128} & Network width & 650 & 64 & 64 \\
                     & Network depth & 8 & 8 & 8 \\
                     & Total trainable parameters & $4.14\times10^{6}$ & $4.32\times10^{6}$ & $1.07\times10^{5}$ \\
\bottomrule
\end{tabular}
\caption{Architectural hyperparameters and total trainable parameters for three Conditional Neural Field variants at latent sizes. “Width” is the number of hidden units per layer; “depth” is the number of hidden layers. CNF-FiLM* denotes the FiLM variant widened to match CNF-FP’s parameter count.}
\label{tab:cnf_params}
\end{table}

\begin{table}[h]
\centering
\begin{tabular}{lcccc}
\toprule
Latent size & Split & CNF-FiLM* & CNF-FP & CNF-FiLM \\
\midrule
\multirow{3}{*}{32}
 & Training      & 0.84\% & 0.46\% & 2.10\% \\
 & Interpolation & 1.67\% & 0.73\% & 2.37\% \\
 & Extrapolation & 6.57\% & 5.21\% & 5.05\% \\
\midrule
\multirow{3}{*}{64}
 & Training      & 0.55\% & 0.39\% & 1.73\% \\
 & Interpolation & 1.59\% & 0.54\% & 2.02\% \\
 & Extrapolation & 5.74\% & 4.37\% & 4.12\% \\
\midrule
\multirow{3}{*}{128}
 & Training      & 0.29\% & 0.21\% & 5.05\% \\
 & Interpolation & 1.06\% & 0.36\% & 1.64\% \\
 & Extrapolation & 4.79\% & 3.61\% & 3.10\% \\
\bottomrule
\end{tabular}
\caption{Reconstruction error (\%) by latent size and data split. CNF-FiLM* is the widened FiLM model with parameter count comparable to CNF-FP at each latent size.}
\label{tab:cnf_errors_equal_param}
\end{table}

% In the main text experiment, we keep the 
\clearpage
\bibliographystyle{elsarticle-num}
\bibliography{submissio-v0/mypub,submissio-v0/ref}
\end{document}